\documentclass[apj]{emulateapj}
\usepackage{xcolor}
\usepackage{amsmath}
\usepackage{amsfonts}
\usepackage{amssymb}
\usepackage{natbib}

\newcommand{\um}{$\mu$m}

\begin{document}

\title{Mass Assembly of Stellar Systems and their Evolution with the SMA (MASSES).  Multiplicity and the Physical Environment in L1448N}
\author{Katherine I.\  Lee\altaffilmark{1}, 
Michael M.\ Dunham\altaffilmark{1}, 
Philip C.\ Myers\altaffilmark{1},  
John J.\ Tobin\altaffilmark{2},
Lars E.\ Kristensen\altaffilmark{1},
Jaime E.\ Pineda\altaffilmark{3},
Eduard I.\ Vorobyov\altaffilmark{4,15},
Stella S.\ R.\ Offner\altaffilmark{5},
H\'{e}ctor G.\ Arce\altaffilmark{6},
Zhi-Yun Li\altaffilmark{7},
Tyler L.\ Bourke\altaffilmark{8,1},
Jes K.\ J\o rgensen\altaffilmark{9},
Alyssa A.\ Goodman\altaffilmark{1},
Sarah I.\ Sadavoy\altaffilmark{10},
Claire J.\ Chandler\altaffilmark{11},
Robert J.\ Harris\altaffilmark{12},
Kaitlin Kratter\altaffilmark{13},
Leslie W.\ Looney\altaffilmark{12},
Carl Melis\altaffilmark{14},
Laura M.\ Perez\altaffilmark{10},
and 
Dominique Segura-Cox\altaffilmark{12},
}
\affil{$^{1}$Harvard-Smithsonian Center for Astrophysics, Cambridge, MA 02138, USA; katherine.lee@cfa.harvard.edu}
\affil{$^{2}$Leiden Observatory, Leiden University, Leiden, The Netherlands}
\affil{$^{3}$Max-Planck-Institut f\"{u}r extraterrestrische Physik, 85748 Garching, Germany}
\affil{$^{4}$Department of Astrophysics, The University of Vienna, Vienna, A-1180, Austria}
\affil{$^{5}$Department of Astronomy, University of Massachusetts, Amherst, MA 01003, USA}
\affil{$^{6}$Department of Astronomy, Yale University, New Haven, CT 06520, USA}
\affil{$^{7}$Department of Astronomy, University of Virginia, Charlottesville, VA 22903, USA}
\affil{$^{8}$SKA Organization, Jodrell Bank Observatory, Lower Withington, Macclesfield, Cheshire SK11 9DL, UK}
\affil{$^{9}$Niels Bohr Institute and Center for Star and Planet Formation, Copenhagen University, DK-1350 Copenhagen K., Denmark}
\affil{$^{10}$Max-Planck-Institut f\"{u}r Astronomie, D-69117 Heidelberg, Germany}
\affil{$^{11}$National Radio Astronomy Observatory, Socorro, NM 87801, USA}
\affil{$^{12}$Department of Astronomy, University of Illinois, Urbana, IL 61801, USA}
\affil{$^{13}$Steward Observatory, University of Arizona, Tucson, AZ 85721, USA}
\affil{$^{14}$Center for Astrophysics and Space Sciences, University of California, San Diego, CA 92093, USA}
\affil{$^{15}$Research Institute of Physics, Southern Federal University, Rostov-on-Don, 344090, Russia}

\begin{abstract}

We present continuum and molecular line observations at 230 GHz and 345 GHz from the Submillimeter Array (SMA) toward three protostars in the Perseus L1448N region.
The data are from the large project ``Mass Assembly of Stellar Systems and their Evolution with the SMA" (MASSES).   
Three dust continuum sources, Source B, Source NW, and Source A, are detected at both frequencies. 
These sources have corresponding emission peaks in C$^{18}$O ($J=2\rightarrow 1$), $^{13}$CO ($J=2\rightarrow 1$), and HCO$^{+}$ ($J=4\rightarrow 3$), and have offsets with N$_{2}$D$^{+}$ ($J = 3\rightarrow 2$) peaks.  
High angular resolution data from a complimentary continuum survey with the Karl G.\ Jansky Very Large Array show that Source B is associated with three 8~mm continuum objects, Source NW with two, and Source A remains single. 
These results suggest that multiplicity in L1448N exists at different spatial scales from a few thousand AU to $<$ 100 AU.  
Velocity gradients in each source obtained from two-dimensional fits to the SMA C$^{18}$O emission are found to be perpendicular to within 20 degrees of the outflow directions as revealed by $^{12}$CO ($J=2\rightarrow 1$). 
We have observed that Sources B and NW with multiplicity have higher densities than Source A without multiplicity.
This suggests that thermal Jeans fragmentation can be relevant in the fragmentation process. 
However, we have not observed a difference in the ratio between rotational and gravitational energy between sources with and without multiplicity. 
We also have not observed a trend between non-thermal velocity dispersions and the level of fragmentation.  
Our study has provided the first direct and comprehensive comparison between multiplicity and core properties in low-mass protostars, although based on small number statistics.  

\end{abstract}

\maketitle

\section{Introduction}

Multiple/binary systems are a common outcome of the star formation process \citep[see reviews by][]{2002ARA&A..40..349T,2014prpl.conf..267R,2013ARA&A..51..269D}.
Approximately 25\%-30\% of M-stars, over 50\% of solar-type stars, and nearly all O-stars are found in multiple systems \citep{2006ApJ...640L..63L,2010ApJS..190....1R,2011IAUS..272..474S}.
A survey of pre-main-sequence stars in Taurus has discovered that nearly half of the sources are binary with separations from 18 AU to 1800 AU \citep{1998A&A...331..977K}.
With interferometric observations, studies have begun to reveal and characterize multiplicity in the protostellar phase in the past two decades \citep[e.g.][]{1997ApJ...484L.157L}.   
For example, \citet{2000ApJ...529..477L} observed $\sim 15$ embedded objects in dust continuum at 2.7~mm and found all the objects are in small groupings or binary systems with most separations ranging from a few hundred AU to a few thousand AU.
In addition, \citet{2013ApJ...768..110C} found 21 out of 33 observed Class 0 objects are in binary/multiple systems with separations ranging from 50 AU to 5000 AU \citep[c.f.,][]{2010A&A...512A..40M}. 
Moreover, with dense gas tracers \citet{2013ApJ...772..100L} found substructures in 7 out of 8 observed starless cores in Orion, which could be seeds for future star formation activities.
Recently, a wide-separation ($>3000$ AU) multiple system in formation was identified in Barnard 5 composed of one protostar and three gravitationally bound gas condensations \citep{2015Natur.518..213P}.
These results strongly suggest that multiplicity occurs at very early stages of star formation.   

Theoretically, fragmentation at the prestellar or protostellar stage is thought to be the main mechanism for multiplicity/binary formation \citep{2002ARA&A..40..349T}.  
There are a few major modes for fragmentation \citep{2007prpl.conf..133G}:
(i) thermal Jeans fragmentation considering only thermal support and gravity, 
(ii) rotational fragmentation during the collapse of a rotating core \citep[e.g.,][]{1979ApJ...234..289B,1993MNRAS.264..798B,2003MNRAS.340...91C}, 
(iii) turbulent fragmentation as a result of density fluctuations in a bound core \citep[e.g.,][]{2010ApJ...725.1485O,2004A&A...423....1J,2004ApJ...600..769F}, 
(iv) disk fragmentation as induced by gravitational instabilities in a disk \citep[e.g.,][]{2010ApJ...708.1585K,2010ApJ...719.1896V,1989ApJ...347..959A}.
Studying gas kinematics in protostellar cores provides an opportunity to distinguish between the scenarios.

However, despite the observational progress in revealing multiplicity at protostellar stages from observations, 
little is known about the connection between multiplicity and natal protostellar core conditions (including physical and kinematic properties): 
are there differences in core properties between protostellar cores forming multiple systems and those forming single systems?  
In addition, the relation between the level of multiplicity and degrees of rotation and turbulence are rarely explored.
A small number of interferometric observations have characterized the rotation and turbulence in various single and multiple systems toward protostars \citep{2007ApJ...669.1058C,2006ApJ...651..301V};
however, no relations between core properties (physical and kinematic) and multiplicity were explicitly and comprehensively investigated in low-mass protostars.   

We are undertaking a large survey with the Submillimeter Array (SMA), ``Mass Assembly of Stellar 
Systems and their Evolution with the SMA" (MASSES; Principle Investigator: Michael M.\ Dunham), 
to address the link between multiplicity and gas environment (physical and kinematic conditions) as one of its primary goals.  MASSES, which 
provides kinematic information on the gas in protostars (see below), is highly 
complementary to the VLA Nascent Disk And Multiplicity (VANDAM) survey 
\citep{2015ApJ...798...61T}, a VLA continuum survey which characterizes 
protostellar multiplicity down to a spatial resolution of 15 AU.  
MASSES targets the same objects as the VANDAM survey: all 73 known protostars 
in the Perseus molecular cloud 
\citep[$d = 230$~pc,][]{2008PASJ...60...37H,2011PASJ...63....1H}, including the 
66 protostars identified by {\it Spitzer} \citep{2009ApJ...692..973E} and 7 
candidate first hydrostatic cores.  By targeting the complete population of 
protostars in a single molecular cloud, MASSES aims to study the origins of 
fragmentation, the evolution of angular momentum in dense star-forming cores, 
and the evolution of molecular outflows, all unified under the common theme of 
developing a more complete understanding of the stellar mass assembly process.  
MASSES targets various molecular lines and continuum observations at both 230 
and 345 GHz in the Subcompact and Extended array configurations, 
providing an angular resolution of $\sim$1$''$ (230 AU at the distance of 
Perseus) while recovering emission up to scales of $\sim$20$''$ (5000 AU at 
the distance of Perseus).  When complete, MASSES will provide the largest, 
unbiased, interferometric sample of protostars observed in the same region 
with uniform sensitivity, angular resolution, and spectral line coverage.

In this paper we present the first results from MASSES by focusing on the 
multiple system L1448N in Perseus in order to study the connection between 
multiplicity and the core kinematics.  
Located in the north of the L1448 complex \citep{1986A&A...168..262B}, 
L1448N (also recognized as L1448 IRS 3) is the brightest source at far-infrared wavelengths among the three \textit{IRAS} sources in L1448 \citep{1986A&A...166..283B}.
Previous single-dish observations estimated core masses of 11.3 M$_{\sun}$ based on 1~mm observations \citep{2006ApJ...638..293E} and 17.3 M$_{\sun}$ based on 850~\um observations \citep{2010ApJ...710.1247S} for L1448N.  
Higher resolution observations of L1448N at millimeter and centimeter wavelengths \citep{1997IAUS..182..507T,2000ApJ...529..477L,2013ApJ...768..110C,1989ApJ...341..208A,1990ApJ...365L..85C,2002AJ....124.1045R} show this source consists of three distinct continuum sources (L1448N-B, L1448N-A, and L1448N-NW). 
All three sources have active outflows and have been suggested to be Class 0 sources \citep[e.g.,][]{2006ApJ...653.1358K,1998ApJ...509..733B,2000ApJS..131..249S}, although L1448N-A could be close to Class I \citep{2006AJ....131.2601O}. 
Recent interferometric continuum observations at 1.3~mm with a 0.3\arcsec \ resolution suggest that L1448N-B harbors a candidate massive protostellar disk \citep{2015ApJ...805..125T}.

\section{Observations}

\begin{deluxetable*}{cccccccc}
\tablewidth{0pc}
\tablecolumns{8}
\tabletypesize{\scriptsize}
\tablecaption{SMA Observation Log}
\tablehead{
\colhead{Config.} & \colhead{Date} & \colhead{Central Frequencies} & \colhead{Tsys} & \colhead{Baseline} & \colhead{Gain} & \colhead{Gaincal Flux} & \colhead{Flux} \\
\colhead{} & \colhead{(UT)} & \colhead{(GHz)} & \colhead{(K)} & \colhead{($k\lambda$)} & \colhead{Calibrator} & \colhead{Density (Jy)} & \colhead{Calibrator}
}
\startdata
Extended & 2014 September 05 & 231.29 & 210 & 20-170 & 3C84 & 11.2 & Neptune \\
Extended & 2014 September 05 & 356.72 & 680 & 28-265 & 3C84 & 7.5 & Neptune \\
Subcompact & 2014 November 18 & 231.29 & 200 & 4-51 & 3C84 & 11.5 & Uranus \\
Subcompact & 2014 November 18 & 356.72 & 550 & 6-82 & 3C84 & 7.5 & Uranus
\enddata
\label{tbl:obs}
\end{deluxetable*}

\begin{deluxetable*}{lcccccccc}
\tablewidth{0pc}
\tablecolumns{10}
\tabletypesize{\scriptsize}
\tablecaption{Summary of Molecular Line and Continuum Data}
\tablehead{
\colhead{Line} & \colhead{Config.} & \colhead{Rest Freq.} & \colhead{Num.} & \colhead{Velo.\ Resolution\tablenotemark{a}} & \colhead{Chan.\ RMS} & \colhead{Synth.\ Beam (P.A.)} & \colhead{Detected?} & \colhead{Fig?\tablenotemark{b}} \\
\colhead{} & \colhead{} & \colhead{(GHz)} & \colhead{of Chan.} & \colhead{(km~s$^{\text{-1}}$)} & \colhead{(mJy~beam$^{\text{-1}}$ (K))} & \colhead{} & \colhead{} & \colhead{}
}
\startdata
$^{12}$CO(2-1) & Extended & 230.53796 & 512 & 0.26/0.5 & 60 (1.14) & 1.26\arcsec$\times$0.86\arcsec (86.5\arcdeg) & Y & \checkmark\\
                & Subcompact &  & &  & 90 (0.15) & 4.23\arcsec$\times$3.24\arcsec (-21.7\arcdeg) & Y & \\
$^{13}$CO(2-1) & Extended & 220.39868 & 512 & 0.26/0.3 & 44 (0.90) & 1.33\arcsec$\times$0.92\arcsec (-90.0\arcdeg) & Y & \\
           & Subcompact &  &  &  & 97 (0.17) & 4.46\arcsec$\times$3.42\arcsec (-20.4\arcdeg) & Y & \\
           & Combined   &  &  &  & 42 (0.40) & 1.79\arcsec$\times$1.51\arcsec (85.44\arcdeg) & Y & \checkmark \\
C$^{18}$O(2-1) & Extended & 219.56036 & 1024 & 0.13/0.2 & 68 (1.41) & 1.33\arcsec$\times$0.92\arcsec (-89.8\arcdeg) & Y & \\
        & Subcompact &  &  &  & 141 (0.25) & 4.35\arcsec$\times$3.35\arcsec (-19.9\arcdeg) & Y & \checkmark \\
N$_{2}$D$^{+}$(3-2) & Extended & 231.32183 & 1024 & 0.13/0.2 & 90 (1.92) & 1.26\arcsec$\times$0.85\arcsec (86.6\arcdeg) & N & \\
            & Subcompact &  &  &  & 145 (0.24) & 4.29\arcsec$\times$3.23\arcsec (-22.4\arcdeg) & Y & \checkmark \\
$^{12}$CO(3-2) & Extended & 345.79599 & 1024 & 0.085/0.5 & 159 (3.41) & 0.85\arcsec$\times$0.56\arcsec (78.8\arcdeg) & Y & \\
           & Subcompact &  &  &  & 378 (0.64) & 3.17\arcsec$\times$1.91\arcsec (-26.3\arcdeg) & Y & \\
HCO$^{+}$(4-3) & Extended & 356.73424 & 1024 & 0.085/0.2 & 285 (6.22) & 0.80\arcsec$\times$0.55\arcsec (74.7\arcdeg) & N & \\
           & Subcompact &  &  &  & 418 (0.67) & 3.16\arcsec$\times$1.93\arcsec (-24.2\arcdeg) & Y & \checkmark \\
H$^{13}$CO$^{+}$(4-3) & Extended & 346.99835 & 1024 & 0.085/0.2 & 221 (4.74) & 0.84\arcsec$\times$0.56\arcsec (78.5\arcdeg) & N & \\
              & Subcompact &  &  &  & 366 (0.61) & 3.26\arcsec$\times$1.92\arcsec (-27.0\arcdeg) & Y & \\
Continuum & Extended & 231.29 & 64 & \nodata & 3.40 & 1.30\arcsec$\times$0.90\arcsec (89.3\arcdeg) & Y & \\
      & Subcompact &  &  &  & 4.15 & 4.27\arcsec$\times$3.26\arcsec (-20.9\arcdeg) & Y & \\
            & Combined & &  &  & 3.66 & 1.87\arcsec$\times$1.68\arcsec (83.53\arcdeg) & Y & \checkmark \\
Continuum & Extended & 356.72 & 64 & \nodata & 5.55 & 0.83\arcsec$\times$0.58\arcsec (79.3\arcdeg) & Y & \\
      & Subcompact &  &  &  & 19.70 & 3.55\arcsec$\times$2.01\arcsec (-26.5\arcdeg) & Y & \\
            & Combined & &  &  & 10.50 & 1.63\arcsec$\times$1.43\arcsec (-10.79\arcdeg) & Y & \checkmark
\enddata
\tablenotetext{a}{The velocity resolutions listed here are the original resolutions (the first numbers) and the resolutions after rebinning (the second numbers).}
\tablenotetext{b}{The datasets that are used in the following figures in this paper.}
\label{tbl:obs2}
\end{deluxetable*}

\subsection{SMA}
The SMA is a submillimeter- and millimeter-wavelength interferometer consisting of eight 6.1~m antennas located on Mauna Kea in Hawaii \citep{2004ApJ...616L...1H}. 
We observed the source, L1448-N, in the Extended and Subcompact configurations in September 2014 and November 2014, respectively.  
We observed simultaneously in the dual receiver mode with the low frequency receiver centered at 231.29 GHz (1.3~mm) and the high frequency receiver centered at 356.72 GHz (850~\um). 
With the Extended configuration all eight antennas were operational; with the Subcompact configuration seven antennas were operational. 
The observations were obtained in good weather conditions, with the zenith opacity at 225 GHz ranging between 0.1 and 0.15 for the Extended configuration and staying at 0.1 for the Subcompact configuration.  
The details of the observations including median system temperatures, and baseline ranges are summarized in Table~\ref{tbl:obs}.

The visibility data were reduced and calibrated using the MIR software package\footnote{Available at https://www.cfa.harvard.edu/$\sim$cqi/mircook.html} following standard calibration procedures.
A baseline correction was first applied to the dataset.
Phases and amplitudes of calibrators on each baseline were then inspected, and data that were not able to be calibrated were manually flagged.
Corrections for system temperatures were applied.
We calibrated bandpass by applying antenna-based solutions derived from 3C84. 
We also used 3C84 for gain calibration by applying antenna-based solutions. 
Flux calibration was performed based on bright quasars or planets, and
the uncertainty in the absolute flux calibration was estimated to be $\sim 25$\%.  
The information of the gain and flux calibrators is summarized in Table~\ref{tbl:obs}.

We observed molecular lines and the continuum with both the low and high frequency receiver.  
For each receiver, the correlator provided 2 GHz bandwidth in each of the lower and upper sidebands.  
Each 2 GHz band has 24 chunks with a useful bandwidth of 82 MHz (due to overlapping channel edges). 
Our correlator setup included 8 chunks (with 64 channels in each chunk) for continuum observations.
The remaining chunks were used for line observations with high spectral resolutions (see Table~\ref{tbl:obs2}).
The continuum was generated by averaging the chunks with 64 channels per chunk and the resulting continuum band has an effective bandwidth of 1312 MHz considering both the upper and lower sidebands.      
The correlator setup is the same for the Extended and Subcompact configurations.

The calibrated visibility data were imaged using the MIRIAD software package \citep{1995ASPC...77..433S}.
All the data were imaged with the parameter ``robust" = 1. 
We re-binned the channels in molecular lines to obtain higher signal-to-noise ratios.  
Table~\ref{tbl:obs2} lists the velocity resolutions after rebinning.

In the Extended configuration, $^{12}$CO ($J=2\rightarrow 1$), $^{13}$CO ($J=2\rightarrow 1$), $^{12}$CO ($J=3\rightarrow 2$), and the continuum are detected; C$^{18}$O ($J=2\rightarrow 1$) is marginally detected.  
In the Subcompact configuration all of the molecular lines and the continuum are detected. 
Table~\ref{tbl:obs2} summarizes the correlator setup, the spectral resolutions, RMS noise levels, and synthesized beams for the observations and calibrated images. 

In this paper we focus on molecular lines, including $^{12}$CO ($J=2\rightarrow 1$), $^{13}$CO ($J=2\rightarrow 1$), C$^{18}$O ($J=2\rightarrow 1$), N$_{2}$D$^{+}$ ($J = 3\rightarrow 2$), and HCO$^{+}$ ($J=4\rightarrow 3$), which show strong detections in the Subcompact and/or Extended data.
We show the Subcompact data for C$^{18}$O ($J=2\rightarrow 1$), N$_{2}$D$^{+}$ ($J = 3\rightarrow 2$), and HCO$^{+}$ ($J=4\rightarrow 3$), since these molecular lines do not have strong detections in the Extended data. 
For $^{13}$CO ($J=2\rightarrow 1$) we show the map combining the Subcompact and Extended data since the molecular line has strong detections from both configurations.    
For $^{12}$CO ($J=2\rightarrow 1$), we use only the extended configuration data to identify outflows because it provides the highest angular resolution and thus the least amount of confusion.
$^{12}$CO ($J=3\rightarrow 2$) was detected in both Subcompact and Extended configurations but is not shown in this paper since it did not provide additional information than $^{12}$CO ($J=2\rightarrow 1$) for the identification of outflows.
Table~\ref{tbl:obs2} summarizes which datasets are being used in the following figures in this paper. 

\subsection{VLA}

The VLA data shown in this paper is from a large survey with the Very Large Array (VLA), the VLA Disk and Multiplicity Survey of Perseus Protostars (VANDAM).  
The observational details of VANDAM are described in \citet{2015ApJ...798...61T}. 
Below we summarize the $Ka-$band observations with the VLA toward L1448N presented in this paper. 

L1448N was observed in the B and A configurations on 2013 November 4 and 2014 February 21, respectively.  
The $Ka-$band observations were conducted with the full continuum mode with one 4 GHz band centered at 36.9 GHz and another 4 GHz band centered at 28.5 GHz. 
The full 8 GHz bandwidth was divided into 128 MHz spectral windows; each window had 64 channels with a channel width of 2 MHz. 
The data were reduced and calibrated using CASA 4.1.0 and version 1.2.2 of the VLA pipeline. 
We used the \textit{clean} task in multi-frequency synthesis mode for imaging.  
The synthesized beam is $0.18\arcsec \times 0.15\arcsec$ (P.A.=70.7\arcdeg).

\section{Results}

\begin{figure*}
\begin{center}
\includegraphics[scale=0.65]{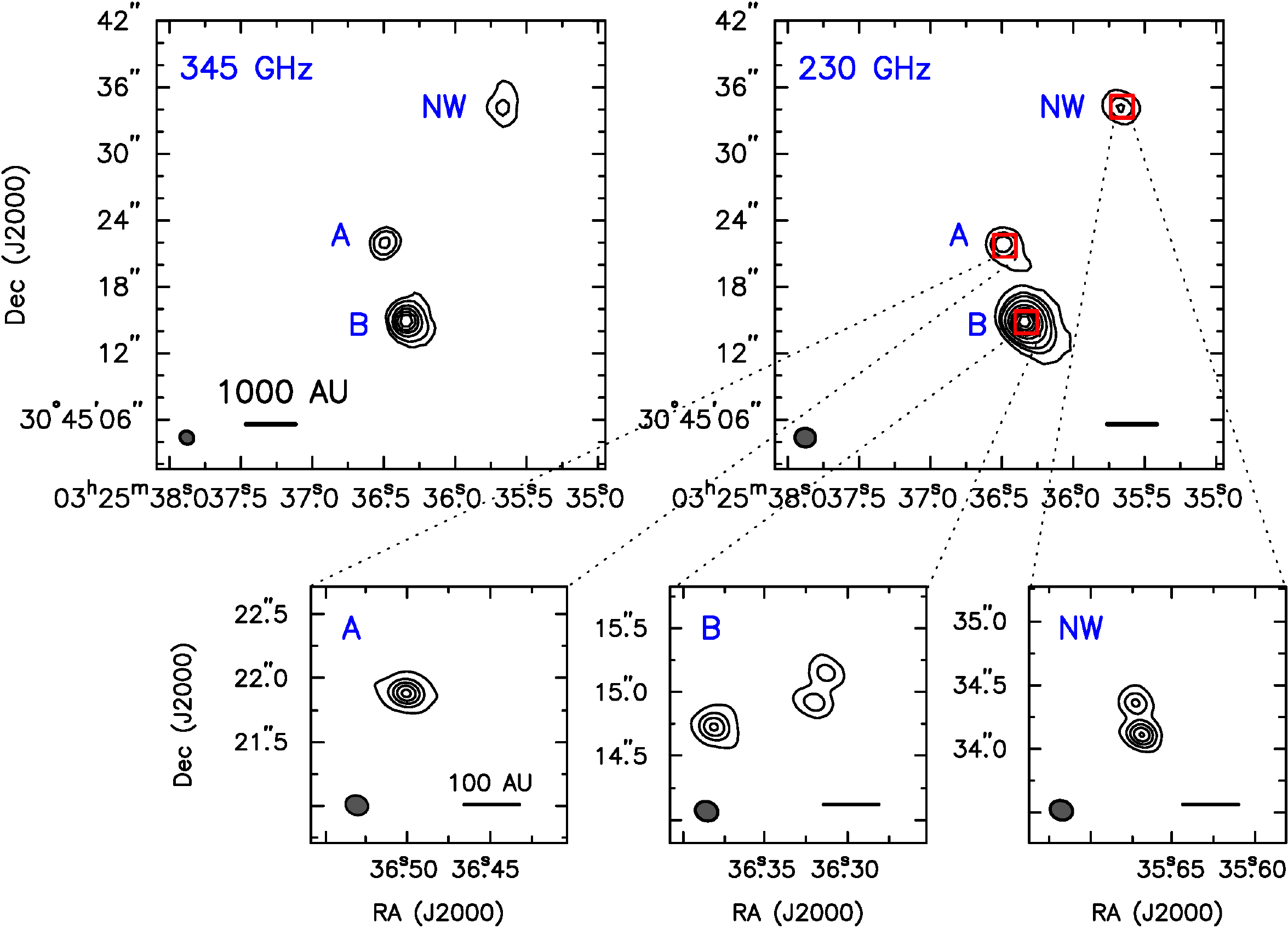}
\caption{
\footnotesize
\textit{Top panels:}
L1448N maps of the 345 GHz (left) and 230 GHz (right) continuum emission.
In the left panel, the contours are 5, 10, 20, 30 ,40 $\times \sigma$ ($\sigma=10.5$ mJy~beam$^{-1}$).  
In the right panel, the contours are 5, 10, 15, 25, 40, 55, 70, 85, 100 $\times \sigma$ ($\sigma=3.66$ mJy~beam$^{-1}$).
Synthesized beams are shown by the filled black ellipses in the lower left corners.
The black solid line shows the scale of 1000 AU. 
The data shown here are from the combination of the Subcompact and Extended data.
\textit{Bottom panels:}
Maps of the 8~mm continuum emission from VANDAM (Tobin et al., submitted).
The contours for Sources B and NW are 10, 25, 40, 55, 70, 85 $\times \sigma$, and the contours for Source A are 10, 40, 70, 100, 130 $\times \sigma$ ($\sigma=0.0056$ mJy~beam$^{-1}$).
Synthesized beams are shown by the filled black ellipses in the lower left corners.
The black solid line shows the scale of 100 AU. 
}
\label{fig:cont}
\end{center}
\end{figure*}

\subsection{Dust Continuum: Multiplicity From 1000 AU to 100 AU Scales}
\label{sect:dust}

\addtolength{\tabcolsep}{-3pt} 
\begin{deluxetable*}{lcccccccccc}
\tablewidth{0pt}
\tablecolumns{11}
\tabletypesize{\scriptsize}
\tablecaption{1.3~mm and 850~\um \ Continuum Properties}
\tablehead{
\colhead{} & \multicolumn{4}{c}{1.3~mm} & \colhead{} & \multicolumn{4}{c}{850~\um} & \colhead{} \\
\cline{2-5} \cline{7-10} \\
\colhead{Source} & \colhead{RA} & \colhead{Dec} & \colhead{Size\tablenotemark{a}} & \colhead{PA\tablenotemark{b}} & \colhead{} & \colhead{RA} & \colhead{Dec} & \colhead{Size\tablenotemark{a}} & \colhead{PA\tablenotemark{b}} & \colhead{Offset\tablenotemark{c}} \\
\colhead{} & \colhead{(J2000)} & \colhead{(J2000)} & \colhead{(maj.\arcsec$\times$min.\arcsec)} & \colhead{(\arcdeg)} & \colhead{} &  \colhead{(J2000)} & \colhead{(J2000)} & \colhead{(maj.\arcsec$\times$min.\arcsec)} & \colhead{(\arcdeg)} & \colhead{(\arcsec)}
}
\startdata
L1448N-B & 03:25:36.33 & +30:45:14.81 & 2.20$\times$1.64 & 35.3 & & 03:25:36.33 & +30:45:14.88 & 1.76$\times$1.49 & 47.4 & (+0.015, +0.062) \\
L1448N-NW & 03:25:35.66 & +30:45:34.26 & 2.27$\times$1.81 & 53.7 & & 03:25:35.65 & +30:45:34.63 & 4.10$\times$2.44 & -2.4 & (-0.125, +0.374) \\
L1448N-A & 03:25:36.48 & +30:45:21.70 & 2.44$\times$1.51 & 34.1 & & 03:25:36.49 & +30:45:21.94 & 1.18$\times$0.89 & -56.6 & (+0.149, +0.241)
\enddata
\label{tbl:cont1}
\tablenotetext{a}{The sizes are deconvolved FWHM sizes (major and minor axes) with the synthesized beam.}
\tablenotetext{b}{Positions angles of major axes from North to East.}
\tablenotetext{c}{Positional offsets between 850~$\micron$ and 1.3~mm}
\end{deluxetable*}

The top panels in Figure~\ref{fig:cont} show the continuum maps at 345 GHz (850~\um) and 230 GHz (1.3~mm).
There are three continuum sources observed, consistent with previous work \citep{2000ApJ...529..477L,2006AJ....131.2601O,2013ApJ...768..110C}.
These three sources are L1448N-B (hereafter Source B), L1448N-A (hereafter Source A), and L1448N-NW (hereafter Source NW).
We fitted a Gaussian to the three sources using the task \textit{imfit} in MIRIAD. 
Table~\ref{tbl:cont1} shows the fitting results including peak positions and FWHM sizes at both 345 GHz and 230 GHz. 
The offsets between the continuum peaks at both frequencies are much smaller than the synthesized beams (Table~\ref{tbl:cont1}), suggesting that the continuum peaks at both frequencies are consistent with each other.

Table~\ref{tbl:cont2} lists the resulting peak intensities and total flux densities from task \textit{imfit} in MIRIAD.
For peak intensities,
Source B has the highest value among the three sources at both 1.3~mm and 850~\um. 
Source NW has a comparable peak intensity to Source A at 1.3~mm, and has a smaller peak intensity at 850~\um \ compared to Source A.  
For total flux densities, 
Source B has the largest values among the three sources at both 1.3~mm and 850~\um, followed by Source NW and then Source A.  
The positions, sizes, peak flux densities, and total flux densities were measured based on primary-beam corrected maps.  
Unless otherwise specified, all calculations are performed using images corrected for the primary beam attenuation.  All images, however, are shown using maps uncorrected for the primary beam attenuation for visual display.

The bottom panels in Fig.~\ref{fig:cont} show the maps of the continuum emission at 8~mm from VANDAM (Tobin et al., submitted).
These results show that Source B and NW are not single systems:
Source B is associated with three 8~mm objects and Source NW with two.
Source A remains single. 
Most of these 8~mm objects are suggested to be protostellar given that their spectral indices are consistent with dust continuum except for the central object in Source B (RA $\sim$ 03$^{\text{h}}$25$^{\text{m}}$38.32$^{\text{s}}$, Dec $\sim$ 30\arcdeg45\arcmin14.85\arcsec), which shows a flat spectral index possibly due to free-free emission (Tobin et al., submitted).
Source B is resolved into at least two objects at 1.3~mm with CARMA \citep{2015ApJ...805..125T}.

The projected separation between Source B and Source A is $\sim 1600$ AU, and that between Source A and Source NW is $\sim 3800$ AU.  
In Source B, the projected separation between the 8~mm object in the East and the two 8~mm objects in the West is $\sim 200$ AU.
In Source NW, the projected separation between the two 8~mm objects is $\sim 50$ AU.  
These results have shown that multiplicity occurs at both a few 1000 AU scales and $50-200$ AU scales in L1448N.
In addition, L1448N is part of the fragmented L1448 system on a few 0.1~pc scales \citep[e.g.,][]{1986A&A...168..262B}. 
Assuming that the multiplicity is due to fragmentation in the prestellar/protostellar stages, these results suggest that the multiplicity in L1448N is consistent with a picture of hierarchical fragmentation where fragmentation takes place at different spatial scales \citep[e.g.,][]{2011ApJ...735...64W,2013ApJ...763...57T}: L1448N fragments into three sources (Sources B, A, and NW) with separations of few thousand AU, and these three sources continue to fragment into smaller objects with separations at $50-200$ AU scales which are observed at 8~mm. 

In the following we use ``L1448N core" to indicate the whole L1448N core which fragments into three continuum ``sources",  and use ``fragments" to indicate the 8~mm objects inside Sources B, A, and NW.   

\begin{deluxetable*}{lcccccccc}
\tablewidth{0pt}
\tablecolumns{9}
\tabletypesize{\scriptsize}
\setlength{\tabcolsep}{0.5in}
\tablecaption{1.3~mm and 850~\um \ Continuum Properties}
\tablehead{
\colhead{} & \multicolumn{2}{c}{Peak Intensity\tablenotemark{a}} & \colhead{} & \multicolumn{2}{c}{Total Flux Density\tablenotemark{a}} & \colhead{} & \multicolumn{2}{c}{Mass\tablenotemark{b}} \\
\cline{2-3} \cline{5-6} \cline{8-9} \\
\colhead{Source} & \colhead{F$_{\rm 1.3mm}$} & \colhead{F$_{\rm 850\micron}$} & \colhead{} & \colhead{F$_{\rm 1.3mm}$} & \colhead{F$_{\rm 850\micron}$} & \colhead{} & \colhead{M$_{\rm 1.3mm}$} & \colhead{M$_{\rm 850\micron}$} \\
\colhead{} & \colhead{(mJy~beam$^{\text{-1}}$)} & \colhead{(mJy~beam$^{\text{-1}}$)} & \colhead{} & \colhead{(mJy)} & \colhead{(mJy)} & \colhead{} & \colhead{(M$_{\sun}$)} & \colhead{(M$_{\sun}$)}}
\startdata
L1448N-B & 423.2$\pm$14.6 & 756.4$\pm$31.1 & & 922.4$\pm$45.0 & 1634.0$\pm$97.0 & & 0.55$\pm$0.02 & 0.23$\pm$0.02 \\
L1448N-NW & 68.1$\pm$5.5  & 209.4$\pm$18.6 & & 158.2$\pm$18.2 & 1131.0$\pm$212.1 & & 0.10$\pm$0.02 & 0.16$\pm$0.02 \\
L1448N-A & 66.6$\pm$7.6   & 274.4$\pm$29.2 & & 149.6$\pm$24.3 & 403.5$\pm$61.0  & & 0.09$\pm$0.02 & 0.06$\pm$0.01
\enddata
\label{tbl:cont2}
\tablenotetext{a}{The uncertainties here are statistical and exclude the 25\% calibration uncertainties.}
\tablenotetext{b}{The uncertainties are estimated based on the uncertainties in the total flux densities and sizes.}
\end{deluxetable*}

\begin{figure*}[ht]
\begin{center}
\includegraphics[scale=1.25]{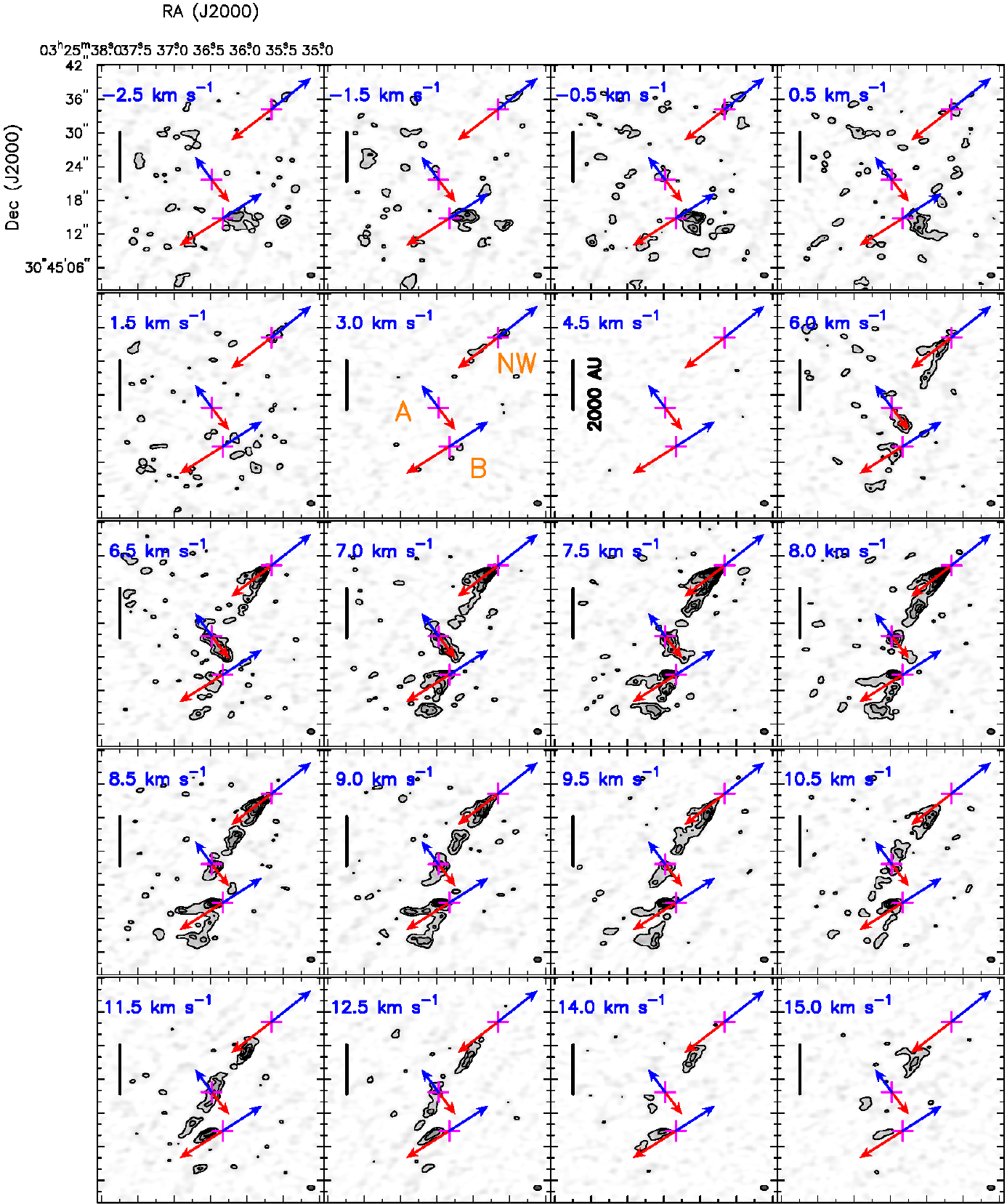}
\caption{
\footnotesize
Channel map of the $^{12}$CO(2-1) emission from the Extended data.
The contours start at 4$\sigma$ and increase with a step of 5$\sigma$ (1$\sigma=$0.06 Jy~beam$^{-1}$).
The three crosses are the locations of Sources B, A, and NW.
The red and blue arrows indicate the directions of the redshifted and blueshifted outflow emission, respectively.
The channel LSR velocity is labeled at the upper left corner in each panel.
The black, solid line indicates the scale of 2000 AU. 
}
\label{fig:outflow}
\end{center}
\end{figure*}

\subsection{$^{12}$CO(2-1): Outflow Directions}

Figure~\ref{fig:outflow} shows the channel maps of $^{12}$CO(2-1) from the Extended Configuration.  
This map provided currently the highest angular resolution in $^{12}$CO(2-1) toward L1448N ($1.26\arcsec \times 0.86\arcsec$).
We detected outflow structures in all three sources (Source B, NW, and A).  
We have identified a cone-like morphology, for the first time, in the red lobe (7.0 - 12.5 km~s$^{-1}$) associated with Source B and our identification of the outflow direction is based on this morphology.
We also acknowledge the possibility that the cone-shaped structure is a line-of-sight overlap of two different outflows stemming from the three 8~mm objects in Source B, where each ``leg" in the cone corresponds to a different outflow. 
In this case one of the two outflow directions is more horizontal than what is currently identified (Fig.~\ref{fig:outflow}), and the more horizontal outflow is consistent with what was identified in \citet{2015ApJ...805..125T}.  

\begin{figure}
\begin{center}
\includegraphics[scale=0.45]{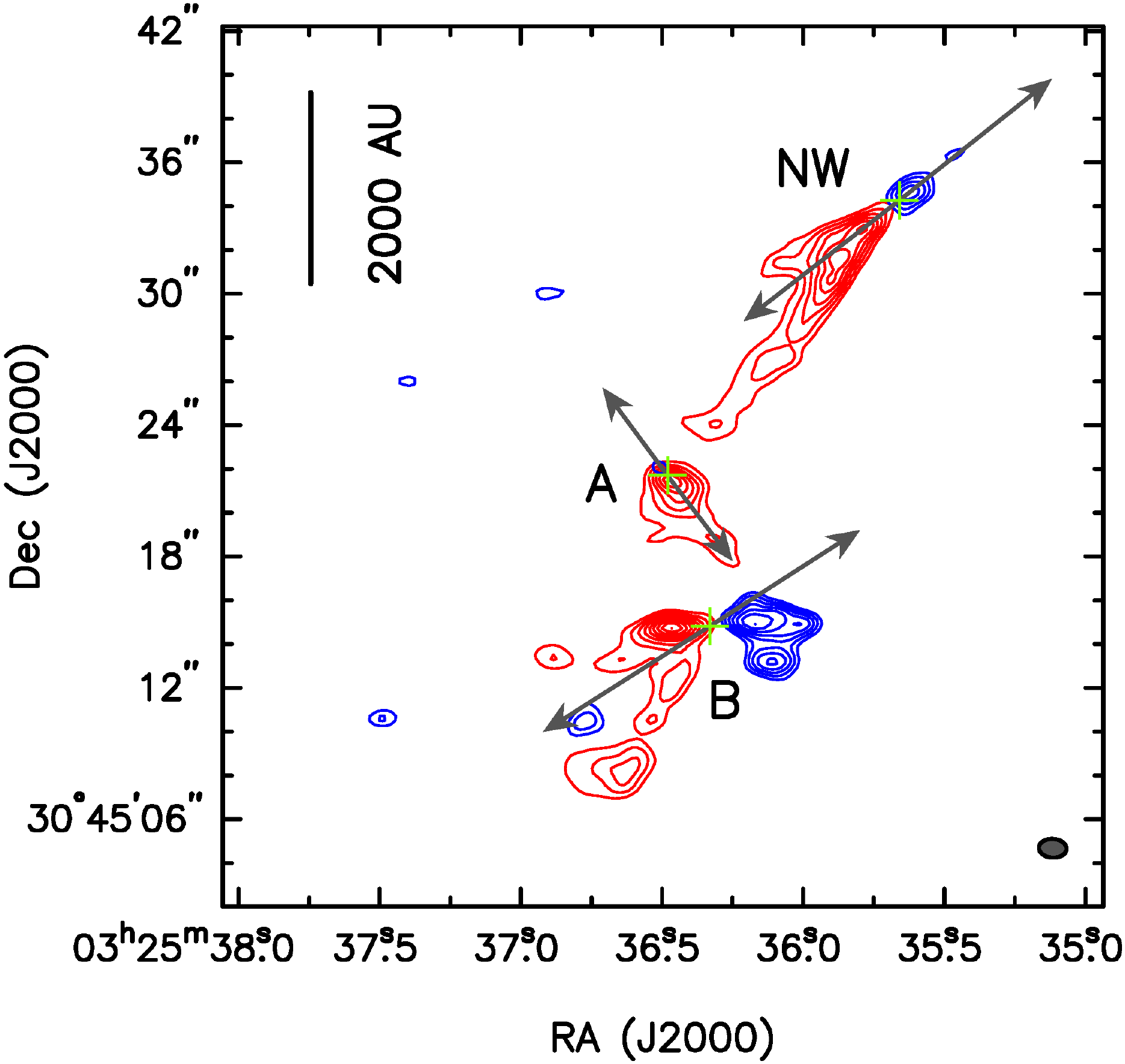}
\caption{
\footnotesize
Integrated intensity maps of the red-shifted and blue-shifted outflows from the $^{12}$CO(2-1) Extended data.
The blue component is integrated from -2.5 km~s$^{-1}$ to 3.0 km~s$^{-1}$ and the red component is integrated from 6.0 km~s$^{-1}$ to 11.0 km~s$^{-1}$.
The red contours are 20\%, 30\%, 40\%, 50\%, 60\%, 70\%, 80\%, 90\%, and 100\% of the peak value (6.5 Jy~beam$^{-1}$~km~s$^{-1}$).
The blue contours are 40\%, 50\%, 60\%, 70\%, 80\%, 90\%, and 100\% of the peak value (3.4 Jy~beam$^{-1}$~km~s$^{-1}$).
The cross symbols show the positions of the three continuum sources.
The grey arrows show the directions of the outflows identified based on the emission morphologies in the channel maps.
}
\label{fig:outflow_int}
\end{center}
\end{figure}

We were also able to identify the outflow direction associated with Source A for the first time.
The redshifted emission appears at channels starting from 6.0 km~s$^{-1}$ up until 9.5 km~s$^{-1}$, beyond which it is contaminated by the outflow from Source NW. 
Source NW is also associated with a cone-like morphology in the redshifted emission (6.5 - 15.0 km~s$^{-1}$), and the identification is consistent with \citet{2015ApJ...805..125T}.
The blueshifted emission is much less visible possibly due to an asymmetric gas distribution in the surrounding environment since Source NW is located near the edge of the whole L1448N core and little dense gas may be farther north. 
The position angles (measured from North to East based on the red-shifted lobes) of the three outflows are 122$\pm 15$ degrees, 218$\pm 10$ degrees, and 128$\pm 15$ degrees for Sources B, A, and NW, respectively, based on manual identification.
Figure~\ref{fig:outflow_int} shows the integrated intensity maps of the red-shifted and blue-shifted outflows with the outflow directions identified based on the channel maps.  

\subsection{Molecular Lines: Morphology and Spectra}
\label{sect:lines}

\begin{figure*}
\begin{center}
\includegraphics[scale=0.83]{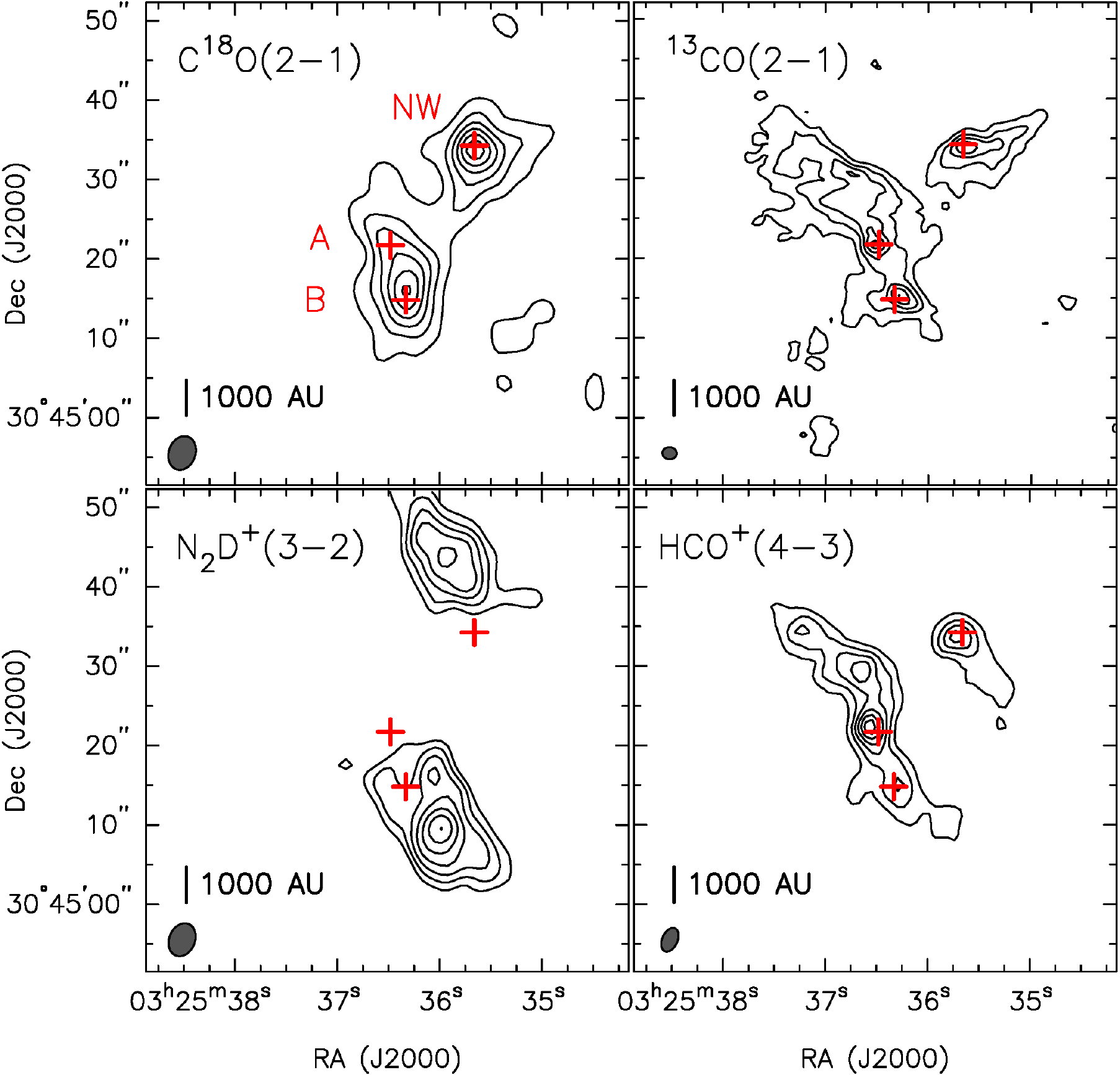}
\caption{
\footnotesize
Integrated-intensity maps of C$^{18}$O(2-1) from the Subcompact data (upper left), $^{13}$CO(2-1) from the combination of the Subcompact and Extended data (upper right), N$_{2}$D$^{+}$(3-2) from the Subcompact data (bottom left), and HCO$^{+}$(4-3) from the Subcompact data (bottom right).
Contours start at $3\sigma$ and increase with a step of $3\sigma$ for C$^{18}$O ($\sigma=0.31$ km~s$^{-1}$~Jy~beam$^{-1}$), $^{13}$CO ($\sigma=0.176$ km~s$^{-1}$~Jy~beam$^{-1}$), HCO$^{+}$(4-3) ($\sigma=0.42$ km~s$^{-1}$~Jy~beam$^{-1}$),
 and a step of $1\sigma$ for N$_{2}$D$^{+}$ ($\sigma=0.29$ km~s$^{-1}$~Jy~beam$^{-1}$).
The red crosses indicate the positions of the three continuum sources B, A, and NW.
Velocity ranges for integration are as follows: (2.2 km~s$^{-1}$ - 6.8 km~s$^{-1}$) for C$^{18}$O, (0.6 km~s$^{-1}$ - 7.5 km~s$^{-1}$) for $^{13}$CO, (3.1 km~s$^{-1}$ - 5.5 km~s$^{-1}$) for N$_{2}$D$^{+}$, and (1.7 km~s$^{-1}$ - 7.7 km~s$^{-1}$) for HCO$^{+}$.
The synthesized beams are shown in the lower left corner of each panel.
}
\label{fig:lines}
\end{center}
\end{figure*}

Figure~\ref{fig:lines} shows the integrated intensity maps of C$^{18}$O(2-1), $^{13}$CO(2-1), N$_{2}$D$^{+}$(3-2), and HCO$^{+}$(4-3).  
All of three continuum sources (shown as red crosses) coincide well with emission peaks in C$^{18}$O, $^{13}$CO, and HCO$^{+}$, suggesting that these molecular lines trace the protostellar envelopes.  
Although C$^{18}$O(2-1) integrated intensities toward Sources A and B are blended, we can see distinct peaks in the channel maps.
The three continuum sources also have corresponding peaks in $^{13}$CO(2-1) and HCO$^{+}$(4-3) with the peak offsets between these two molecules less than one synthesized beam FWHM.  
On the other hand, the N$_{2}$D$^{+}$ peaks do not show correspondence with the continuum sources. 
There are clear offsets between the continuum sources and nearby N$_{2}$D$^{+}$ emission with separations of at least one beam. 

Lines from CO isotopologues including C$^{18}$O and $^{13}$CO have been extensively used to probe protostellar envelopes \citep[e.g.,][]{2007ApJ...659..479J,2015ApJ...799..193Y} and embedded disks \citep{2012Natur.492...83T,2013A&A...560A.103M,2014ApJ...796..131O}.
They are abundant in protostellar structures and are optically thin compared to $^{12}$CO.
In addition, 
while CO has been shown to freeze-out onto cold dust grains in the prestellar/starless core stage \citep{2002ApJ...569..815T, 2013A&A...560A..41L}, 
it is released back to the gas phase in regions above the CO evaporation temperature (20-30 K) in regions ranging from a few hundred to a few thousand AU \citep{2002A&A...389..908J, 2010A&A...518A..52A, 2012ApJ...760...40A, 2012A&A...542A..86Y,2015A&A...579A..23J}. 
As the CO abundance increases, N$_{2}$D$^{+}$ is destroyed by CO and prevented from reforming because H$_{2}$D$^{+}$ is less abundant in regions with T $>$20 K \citep{2009A&A...493...89E,2013ApJ...765...18T}. 
This likely causes the observed offsets between N$_{2}$D$^{+}$ and the CO isotoplogues shown in Fig.~\ref{fig:lines}.   

Emission extending toward the north-east direction from Source A is observed in both HCO$^{+}$ and $^{13}$CO. 
The emission shows clumpiness in HCO$^{+}$ and is more extended in $^{13}$CO.  
As the critical density of the HCO$^{+}$(4-3) transition is $\sim 3\times 10^{6}$~cm$^{-3}$ \citep[e.g.,][]{2015PASP..127..299S} and these HCO$^{+}$ clumps do not have corresponding dust continuum emission at millimeter-wavelengths or infrared, they are likely tracing dense gas in starless fragments.
The reason why C$^{18}$O does not show corresponding emission in these regions is likely to be due to lower column densities and thus is not detected with the sensitivity of the instrument.
Depletion of C$^{18}$O due to freeze-out in the interior of these starless fragments \citep{2004A&A...416..191T,2011ApJ...728..144F} could also contribute to the low column densities.  
These starless fragments demonstrate that L1448N is a system which harbors younger sources in addition to protostars.   
Furthermore, it is noteworthy that Sources A, B, and the HCO$^{+}$ starless fragments appear to be forming on a filamentary structure, an active mode for the star formation process \citep[e.g.,][]{2012ApJ...761..171L}.  
Future observations with better sensitivity are needed to obtain a total mass measurement of these starless fragments to characterize their boundedness and potential for forming future protostars.

\begin{figure*}
\begin{center}
\includegraphics[scale=0.8]{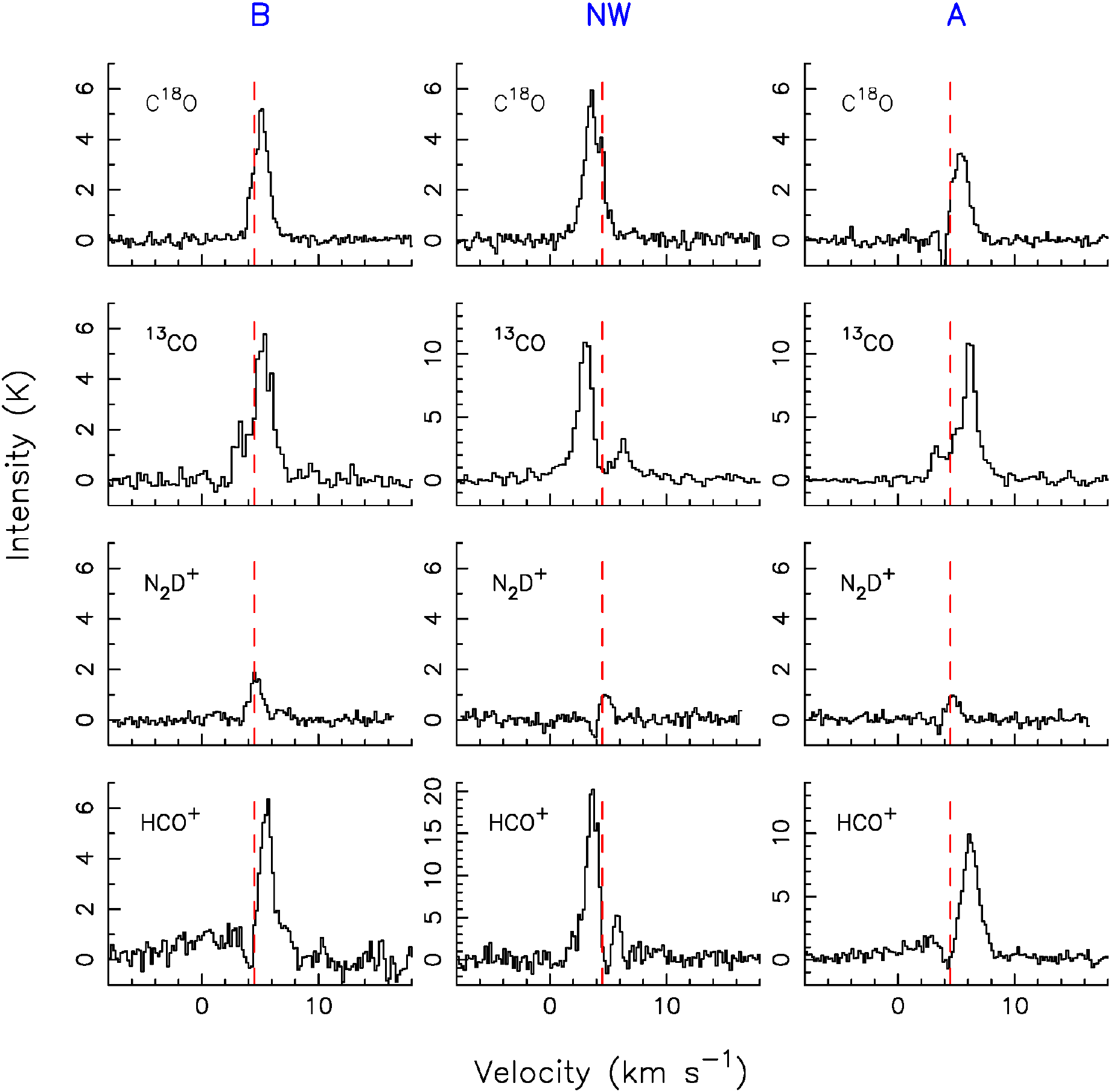}
\caption{
\footnotesize
    Spectra at the position of the continuum sources.
The spectra of C$^{18}$O(2-1) (first row), $^{13}$CO(2-1) (second row), N$_{2}$D$^{+}$(3-2) (third row), and HCO$^{+}$ (bottom row) for Sources B (first panel), NW (second panel), and A (the third panel).
Each spectrum was obtained by averaging the intensities over one beam area at the position of the continuum source.
The red, dashed line indicates the core velocity, 4.5 km~s$^{-1}$, measured from averaging peak velocities over a 31\arcsec \ beam \citep{2008ApJS..175..509R}.
}
\label{fig:spec}
\end{center}
\end{figure*}

Figure~\ref{fig:spec} shows the average spectra at the peak intensities of the four molecular lines in Sources B, NW, and A.
These were obtained by averaging the spectra contained in one beam that was centered on the continuum peaks.
Sources B and NW show similar peak intensities in C$^{18}$O ($\sim$ 5.5 K), whereas Source A is weaker ($\sim$ 3.8 K).
It is also seen that Sources A and B are redshifted compared to the averaged core velocity shown as red, dashed line \citep{2008ApJS..175..509R}, and Source NW is blueshifted.   
For $^{13}$CO, Source NW and A show similar strengths ($\sim$ 11 K), whereas Source B has a lower intensity ($\sim$ 5.8 K).
The peak velocities are consistent with the C$^{18}$O peak velocities. 
Emission at higher velocities is observed in all three sources, likely caused by outflows.  
For N$_{2}$D$^{+}$, emission is weak toward the three sources.
For HCO$^{+}$, Source NW is the strongest among the three sources ($\sim$ 20 K), and Source B is the weakest ($\sim$ 6 K). 
An asymmetric line profile with a stronger blue component and a weaker red component which resembles infall signatures \citep{1999ARA&A..37..311E} is observed in Source NW. 
However, C$^{18}$O, an optically thin line, does not peak where HCO$^{+}$ dips.  
The peak velocity in the blue component of HCO$^{+}$ is consistent with the peak velocity in C$^{18}$O, suggesting that this blue profile is not due to infall.
Also, the dip in the HCO$^{+}$ spectra could be caused by missing flux from the interferometer. 
The red component is possibly due to outflows as emission at the same velocities is observed also in $^{13}$CO.

\section{Mass Estimation}

To characterize the gas masses of the three sources we used the emission from (1) dust continuum and (2) C$^{18}$O(2-1).
As shown in Fig.~\ref{fig:lines}, the C$^{18}$O emission peaks coincide with the dust continuum peaks, suggesting that C$^{18}$O traces high-column density gas in the protostellar envelopes (see also Sect.~\ref{sect:lines}).
Since the C$^{18}$O peaks and dust continuum peaks coincide well, we also use ``Source B", ``Source NW", and ``Source A" to refer to the three C$^{18}$O sources. 

\subsection{Mass Estimates From C$^{18}$O(2-1) Emission}

As Fig.~\ref{fig:lines} shows, Source B and Source A are blended together in the C$^{18}$O integrated intensity map, and therefore we fit two Gaussians to the blended structure to estimate the C$^{18}$O emission associated with each protostellar core. 
The fitting was performed using the \textit{imfit} task in MIRIAD.
Table~\ref{tbl:c18o} lists the fitting results including RA, Dec, FWHM major and minor axes, and position angles. 
We also estimated the effective radius of the sources to be half of the geometric mean of the major and minor axis diameters ($R = \frac{1}{2}\sqrt{\rm maj. \times min.}$).  
The three sources have comparable effective radii ranging from $\sim 800$ AU to $\sim 900$ AU.  
Table~\ref{tbl:c18o} also lists the peak intensities and total flux densities of the sources from the fitting. 
Source NW has the largest values in both peak intensity and total flux density. 
Source B is comparable to Source NW, and Source A has the smallest values in both.  

We first calculated gas column densities by assuming that the C$^{18}$O emission is optically thin and LTE conditions \citep[e.g.,][]{2014ApJ...783...29D}:  
$$
N_{\rm H_{2}} = \dfrac{1}{X_{\rm C^{18}O}} \dfrac{3k}{8\pi\nu \mu^{2}} \dfrac{(2J+1)}{(J+1)} \dfrac{Q(T)}{g_{J}} e^{\frac{E_{J+1}}{kT}} \int T_{\rm mb} \ dV, 
$$
where $X_{\rm C^{18}O}$ is the abundance ratio between C$^{18}$O and molecular hydrogen ([C$^{18}$O]/[H$_{2}$]), $k$ is the Boltzmann constant, $\nu$ is the rest frequency of the transition, $\mu$ is the dipole moment of the molecule, $J$ is the lower state in the molecular transition, $Q(T)$ is the partition function with a rotational temperature T, $g_{J}$ is the statistical weight in the lower state, $E_{J+1}$ is the energy level in the upper state, $\int T_{\rm mb}\ dV$ is the integrated intensity measured in K~km~s$^{-1}$. 
As most of the 15 Class 0 sources have rotational temperatures between 32 K to 39 K from \citet{2013A&A...556A..89Y} (derived from CO lines),
we used 36 K for the rotational temperature.
For $X_{\rm C^{18}O}$, we used $5.2\times 10^{-8}$ as an averaged value from the inner envelope inside the evaporation temperature and the drop-zone abundance in the freeze-out region \citep{2015A&A...576A.109Y}.   
The gas mass of molecular hydrogen was then estimated by integrating under the areas with fitted FWHM major and minor axes; we used a mean molecular weight of 2.8 \citep{2008A&A...487..993K}.

The gas masses are summarized in Table~\ref{tbl:c18o}.
Sources NW and B have comparable masses: $\sim 0.28$ M$_{\sun}$ and $\sim 0.24$ M$_{\sun}$, respectively.  
Source A has $\sim 0.09$ M$_{\sun}$.  
These masses estimated from C$^{18}$O emission are consistent with the mass estimates based on dust continuum emission at 850~\um \ (Table~\ref{tbl:cont2}). 

\subsection{Mass Estimates From Dust Continuum Emission}

\begin{deluxetable*}{lcccccccc}
\tablecolumns{9}
\tabletypesize{\scriptsize}
\tablecaption{C$^{18}$O(2-1) Emission Properties}
\tablehead{
\colhead{Source} & \colhead{RA} & \colhead{Dec} & \colhead{Size\tablenotemark{a}} & \colhead{P.A.} & \colhead{R\tablenotemark{b}} & \colhead{Peak Intensity\tablenotemark{c}} & \colhead{Total Flux Density\tablenotemark{c}} & \colhead{Mass\tablenotemark{d}} \\
\colhead{} & \colhead{(J2000)} & \colhead{(J2000)} & \colhead{(maj.\arcsec$\times$min.\arcsec)} & \colhead{(\arcdeg)} & \colhead{(AU)} & \colhead{(Jy~beam$^{-1}$~km~s$^{-1}$)} & \colhead{(Jy~km~s$^{-1}$)} & \colhead{(M$_{\sun}$)}
}
\startdata
L1448N-B & 03:25:36.325 & +30:45:16.16 & 9.93$\times$5.86 & -5.8 & 878$\pm$123 & 5.25$\pm$0.51 & 28.5$\pm$5.3 & 0.24$\pm$0.04 \\
L1448N-NW & 03:25:35.622 & +30:45:33.16 & 7.45$\times$6.59 & 3.4 & 806$\pm$105 & 6.39$\pm$0.58 & 31.02$\pm$4.9 & 0.28$\pm$0.04 \\
L1448N-A & 03:25:36.573 & +30:45:24.90 & 8.83$\times$7.00 & 2.6 & 904$\pm$334 & 1.81$\pm$0.51 & 9.88$\pm$4.5 & 0.09$\pm$0.04
\enddata
\label{tbl:c18o}
\tablenotetext{a}{The sizes are deconvolved FWHM sizes with the synthesized beam.}
\tablenotetext{b}{R = $\frac{1}{2} \sqrt{\rm major \times \rm minor}$}
\tablenotetext{c}{The uncertainties are statistical from fitting, not systematic.}
\tablenotetext{d}{The uncertainties are estimated based on the uncertainties in the total flux densities and sizes.}
\end{deluxetable*}

Assuming that dust continuum emission is optically thin at both 850~\um \ and 1.3~mm, 
the total gas mass can be estimated using the standard formula:  
\mbox{$M= \dfrac{F_{\nu} D^{2}}{\kappa_{\nu}B_{\nu}(T_{d})}$}, where $F_{\nu}$ is the total flux density of the source, D is the distance to the source, $\kappa_{\nu}$ is the dust opacity at the observed frequency, $B_{\nu}$ is the Planck function, and $T_{d}$ is the dust temperature.    
Here we adopted specific formulas for 850~\um \ and 1.3~mm based on the formula above from \citet{2007ApJ...659..479J} as the following:

\scriptsize
$$
M_{\rm 1.3mm} = 1.3\ M_{\sun} \Big(\dfrac{F_{\rm 1.3mm}}{1\ \text{Jy}}\Big) \Big(\dfrac{D}{200\ \text{pc}}\Big)^{2} \times \Big\{ \text{exp}\Big[0.36\ \Big(\dfrac{30\ K}{T_{d}}\Big) \Big] - 1 \Big\},
$$

$$
M_{\rm 0.85mm} = 0.18\ M_{\sun} \Big(\dfrac{F_{\rm 0.85mm}}{1\ \text{Jy}}\Big) \Big(\dfrac{D}{200\ \text{pc}}\Big)^{2} \times \Big\{ \text{exp}\Big[0.55\ \Big(\dfrac{30\ K}{T_{d}}\Big) \Big] - 1 \Big\}. 
$$
\normalsize

These formulas use dust opacities from \citet{1994A&A...291..943O} based on the models with thin ice mantles coagulated at 10$^{6}$~cm$^{-3}$.  
The gas-to-dust ratio is assumed to be 100.  
We assumed a dust temperature of 36 K for all the three sources, the same as the gas rotational temperature.

Table~\ref{tbl:cont2} summarizes the masses from dust continuum emission at 1.3~mm and 850~\um. 
Based on the emission at 1.3~mm, Source B is $\sim 0.6$ M$_{\sun}$, and Sources NW and A have comparable masses of $\sim0.1$ M$_{\sun}$.  
Based on the emission at 850~\um, Source B is $\sim 0.23$ M$_{\sun}$, Source NW is $\sim 0.16$ M$_{\sun}$, and Source A is $\sim 0.06$ M$_{\sun}$.
In both estimates Source B is the most massive protostellar core among the three sources.
The difference in the masses between the two frequencies could be due to spatial filtering by the interferometry.  
Also, the uncertainties of the estimates can easily be a factor of a few (or more) due to the choices of dust opacities, uncertainties in the total flux densities, and temperature \citep[e.g.,][]{2014MNRAS.444..887D}.

\section{Kinematic Structures in L1448N}

To compare the level of fragmentation revealed by VANDAM (Sect.~\ref{sect:dust}) and kinematics in the three sources, we characterized kinematic motions in terms of velocity gradients and velocity dispersions.
Below we discuss velocity gradients and velocity dispersions separately. 

\subsection{Velocity Gradients}
\label{sect:vgrad}

\begin{figure*}
\begin{center}
\includegraphics[scale=0.81]{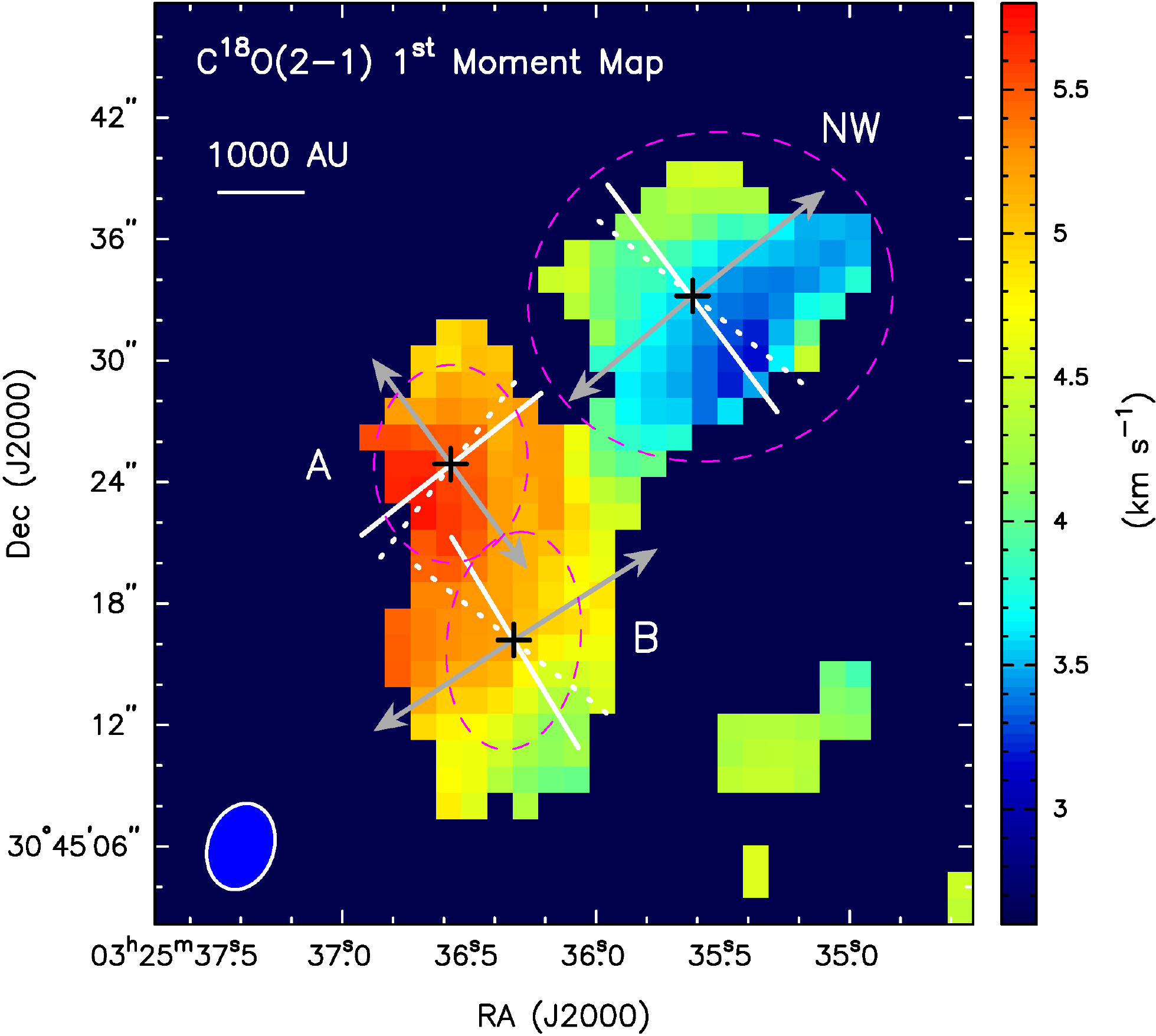}
\caption{
\footnotesize
C$^{18}$O(2-1) 1$^{\rm st}$ moment map, showing the velocity structure of the region.
Pixels below $3\sigma$ in the integrated intensity map are masked.
The grey arrows indicate the directions of the outflows (Fig.~\ref{fig:outflow}).
The white, dashed lines indicate the directions of 2D velocity gradients.
The white, solid lines indicate the directions perpendicular to the outflows.
The black crosses are the positions of the fitted centers from the C$^{18}$O integrated intensity map (Table~\ref{tbl:c18o}).
The magenta, dashed lines show the areas used for 2D velocity gradient fitting.
For Sources B and A, the areas are the FWHM sizes.  For Source NW we used the whole core.
}
\label{fig:mom1}
\end{center}
\end{figure*}

Most of the C$^{18}$O(2-1) spectra are singly peaked, allowing us to investigate gas velocities and dispersions directly using the $1^{\rm st}$ and $2^{\rm nd}$ moment maps.  
We re-gridded the $1^{\rm st}$ and $2^{\rm nd}$ moment maps from slightly oversampled channel maps (0.8\arcsec \ per pixel) with a pixel size of 1.3\arcsec \ based on the Nyquist sampling rate.  
All analysis based on these moment maps is performed on maps with Nyquist sampled pixels to avoid correlations between neighboring pixels.

Figure~\ref{fig:mom1} shows the $1^{\rm st}$ moment map; the positions of fitted C$^{18}$O peaks are marked as black crosses.
All three sources are associated with obvious velocity gradients. 
Using the $1^{\rm st}$ moment map, we computed the magnitudes and/or orientation of the velocity gradients using two methods. 
First, we computed velocity gradients along the direction perpendicular to the outflows (white, solid lines in Fig.~\ref{fig:mom1}) by applying one-dimensional linear fitting.  
These velocity gradients ($\nabla V_{\rm 1D}$) are corrected for inclination, which is measured as 63$\arcdeg$ for Source B based on both images of scattered light at infrared wavelengths and SED fits \citep{2007ApJ...659.1404T}. 
We assume the same inclination angle for all three sources due to the lack of information about the inclination angles of Source A and Source NW from the Literature.

Second, we calculated the magnitudes and orientations of velocity gradients based on two-dimensional linear fitting.
We used the same fitting method as described in \citet{1993ApJ...406..528G}, and the fitting was performed using the IDL \textit{MPFIT} package.
For Sources B and A, we used the fitted Gaussian FWHMs from the C$^{18}$O integrated intensity (shown as magenta, dashed lines in Fig.~\ref{fig:mom1}) as the fitting areas to separate the blended structures. 
For Source NW we used the whole area as indicated by the magenta line. 
Table~\ref{tbl:kinematics} lists the magnitudes of velocity gradients from both 1D fitting ($\nabla V_{\rm 1D}$) and 2D fitting ($\nabla V_{\rm 2D}$) as well as the orientations of the velocity gradients from the 2D fitting ($\theta_{\rm 2D}$).

The velocity gradients from both 1D and 2D fittings are generally consistent for all the three sources. 
Source B has the largest velocity gradient ($\sim 113.6$ km~s$^{-1}$~pc$^{-1}$ from 1D and $\sim 133.8$ km~s$^{-1}$~pc$^{-1}$ from 2D), followed by Source NW ($\sim 89.5$ km~s$^{-1}$~pc$^{-1}$ from 1D and $\sim 76.4$ km~s$^{-1}$~pc$^{-1}$ from 2D), and the smallest is Source A ($\sim 67.5$ km~s$^{-1}$~pc$^{-1}$ from 1D and $\sim 71.5$ km~s$^{-1}$~pc$^{-1}$ from 2D). 
These velocity gradients range from $\sim 70$ km~s$^{-1}$~pc$^{-1}$ to $\sim 130$ km~s$^{-1}$~pc$^{-1}$. 
Our magnitudes are generally consistent with \citet{2015ApJ...799..193Y} as the study reported 183 km~s$^{-1}$~pc$^{-1}$ for Source B.  

These magnitudes are significantly larger than those found in NH$_{3}$/N$_{2}$H$^{+}$ dense cores at 0.1~pc with a typical range of 0.1 - 3.5 km~s$^{-1}$~pc$^{-1}$ \citep{1993ApJ...406..528G,2002ApJ...572..238C}. 
Indeed, velocity gradients in protostellar cores measured with recent observations have shown a wide range of magnitudes \citep[e.g.,][]{2007ApJ...669.1058C,2010MNRAS.408.1516C,2013EAS....62...25B}.
For example, \citet{2011ApJ...740...45T} observed 17 protostellar cores with single-dish telescopes and interferometers using N$_{2}$H$^{+}$(1-0) and NH$_{3}$, and found velocity gradients ranging from $\sim 1$ km~s$^{-1}$~pc$^{-1}$ to $\sim 10$ km~s$^{-1}$~pc$^{-1}$ over a few hundred to a few thousand AU. 
In addition, \citet{2011ApJ...743..201P} measured approximately 6 km~s$^{-1}$~pc$^{-1}$ on a few thousand AU scale toward L1451-mm, a candidate for a first hydrostatic core.  
More recently, \citet{2015ApJ...799..193Y} observed 17 protostellar cores in C$^{18}$O with interferometers and discovered velocity gradients from 1 km~s$^{-1}$~pc$^{-1}$ to 530 km~s$^{-1}$~pc$^{-1}$ with a median value of $\sim 71.6$ km~s$^{-1}$~pc$^{-1}$ over a scale of few thousand AU.

These studies have demonstrated that velocity gradients increase significantly from large (0.1~pc) to small (few thousand AU) scales.  
If the velocity gradients are due to rotation, the significant increase in velocity gradients could suggest that protostellar envelopes rotate faster when going from large to small scales, consistent with the conservation of angular momentum.   
 
We also estimated the differences in the orientation between the 2D velocity gradients and the axis perpendicular to the outflows ($\Delta \theta$ in Table~\ref{tbl:kinematics}).
The two axes are consistent within 20$\arcdeg$, suggesting that the 2D velocity gradients are nearly perpendicular to the outflow directions.
Velocity gradients perpendicular to outflow directions are not uncommon: 
\citet{2011ApJ...740...45T} reported 12 out of the 14 protostars observed with interferometers have velocity gradients normal to outflows within 45$\arcdeg$, and \citet{2015ApJ...799..193Y} found 7 out of 17 observed protostars have velocity gradients close to perpendicular to their outflows.
The angle between the velocity gradients and the outflow directions could change as a source evolves, a scenario proposed by \citet{2006ApJ...646.1070A}.
When complete, MASSES will allow us to fully test this scenario with an unbiased sample which covers the full evolutionary spectrum of protostars.  

\subsection{Velocity Dispersion}

\begin{figure}
\begin{center}
\includegraphics[scale=0.46]{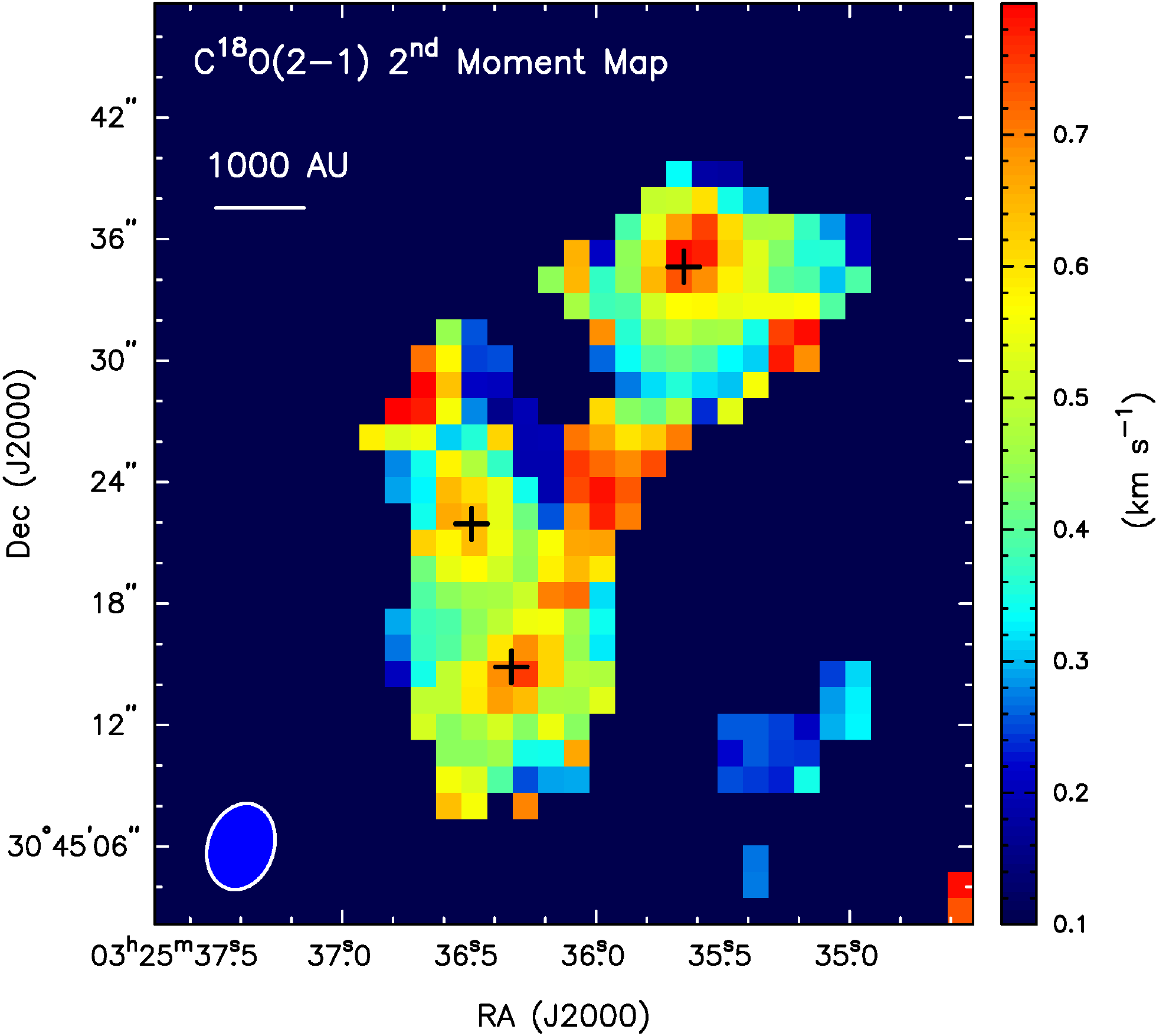}
\caption{
\footnotesize
The $2^{\rm nd}$ moment map of C$^{18}$O(2-1) indicating velocity dispersions.
The black crosses are the positions of the dust continuum sources.
}
\label{fig:mom2}
\end{center}
\end{figure}

Figure~\ref{fig:mom2} shows the $2^{\rm nd}$ moment map from C$^{18}$O indicating velocity dispersions (velocity dispersion $\sigma=$ FWHM$/2.35$ if a profile was a Gaussian).  
For all three sources, their velocity dispersions appear to increase toward the centers of the sources and peak near the positions of the continuum peaks (black crosses).  
We estimated the total velocity dispersion of the whole protostellar core by applying Gaussian fitting to the averaged spectra over the protostellar cores (magenta lines in Fig.~\ref{fig:mom1}).  
Figure~\ref{fig:vdisp} shows the spectra of the three sources averaged over the areas enclosed by the magenta lines in Fig.~\ref{fig:mom1}.  
The fitted peak velocities ($V_{c}$) and velocity dispersions ($\sigma_{\rm tot}$) are listed in Table~\ref{tbl:kinematics}. 
The total velocity dispersions are $\sim$0.66 km~s$^{-1}$, $\sim$0.71 km~s$^{-1}$, and $\sim$0.50 km~s$^{-1}$ for Sources B, NW, and A, respectively.
These dispersions have non-thermal and thermal components.

One of the main contributors to non-thermal velocity dispersions is the shifts in peak velocities, which result in the observed velocity gradients.
To remove this contribution from the total velocity dispersions, 
for each source we estimated the broadening in the dispersions due to velocity shifts over the whole sources based on the 1D fit to the velocity gradient: 
$\delta V_{R} = \nabla V_{\rm 1D} \times R$, where R is the effective radius of the source (Table~\ref{tbl:c18o}).  
We then subtracted $\delta V_{R}$ from $\sigma_{\rm tot}$ and approximated the velocity dispersions without contributions from velocity shifts ($\sigma_{\rm norot}$): $\sigma_{\rm norot}^{2} = \sigma_{\rm tot}^{2} - \delta V_{R}^{2}$, as $\delta V_{R}$ has the biggest contribution to broadening caused by velocity shifts.  

Next, we removed the contribution from the thermal dispersion ($\sigma_{T}$): $\sigma_{T} = \sqrt{(\frac{kT}{\mu m_{\rm H}})}$, where $k$ is the Boltzmann constant, T is the gas kinetic temperature, $\mu$ is the molecular weight of the observed molecular ($\mu = 30$ for C$^{18}$O), and $m_{\rm H}$ is the mass of molecular hydrogen.
The thermal dispersion is 0.091 km~s$^{-1}$ assuming T = 36 K. 
The non-thermal velocity dispersions ($\sigma_{\rm NT}$) after removing the thermal dispersion ($\sigma_{\rm NT}^{2} = \sigma_{\rm norot}^{2} - \sigma_{T}^{2}$) are 0.44 km~s$^{-1}$, 0.62 km~s$^{-1}$, and 0.39 km~s$^{-1}$ for Source B, NW, and A, respectively.   
The results of $\sigma_{\rm tot}$, $\sigma_{\rm norot}$, and $\sigma_{\rm NT}$ are listed in Table~\ref{tbl:kinematics}. 

The remaining non-thermal velocity dispersions can have contributions from motions including turbulence, infall, outflows close to the source, and unresolved rotation at source centers due to disks.
Determining the exact contribution of each component requires detailed modeling.
Here we assume that the non-thermal component of the remaining velocity dispersion is mainly due to turbulence.
As the sound speed is 0.3 km~s$^{-1}$ at 36 K (using a mean molecular weight of 2.8),  
these non-thermal dispersions are transonic (Source A) to supersonic (Source NW), suggesting that turbulent motions could provide significant support, even after accounting for simple rotation.   

\begin{deluxetable*}{ccccccccccc}
\tablecolumns{11}
\tabletypesize{\scriptsize}
\tablecaption{Kinematic Properties \label{tbl:kinematics}}
\tablehead{
\colhead{Source} & \colhead{$\nabla V_{\rm 1D}$\tablenotemark{a}} & \colhead{$\nabla V_{\rm 2D}$\tablenotemark{b}} & \colhead{$\theta_{\rm 2D}$\tablenotemark{c}} & \colhead{$\Delta \theta$\tablenotemark{d}} & \colhead{$\delta V_{R}$\tablenotemark{e}} & \colhead{$\beta_{\rm rot}$\tablenotemark{f}} & \colhead{$V_{c}$\tablenotemark{g}} & \colhead{$\sigma_{\rm tot}$\tablenotemark{h}} & \colhead{$\sigma_{\rm norot}$\tablenotemark{i}} & \colhead{$\sigma_{\rm NT}$\tablenotemark{j}} \\
\colhead{} & \multicolumn{2}{c}{(km~s$^{-1}$~pc$^{-1}$)} & \colhead{($\arcdeg$)} & \colhead{($\arcdeg$)} & \colhead{(km~s$^{-1}$)} & \colhead{} & \colhead{(km~s$^{-1}$)} & \colhead{(km~s$^{-1}$)} & \colhead{(km~s$^{-1}$)} & \colhead{(km~s$^{-1}$)}
}
\startdata
L1448N-B  & $113.6\pm10.0$ & $133.8\pm7.2$ & $32\pm3$ & $21\pm15$ & $0.48\pm0.04$ & $0.10\pm0.05$ & $5.05\pm0.01$ & $0.66\pm0.01$  & $0.45\pm0.04$ & $0.44\pm0.04$ \\
L1448N-NW & $89.5\pm16.5$  & $76.4\pm 5.3$ & $127\pm4$ & $14\pm16$ & $0.35\pm0.06$ & $0.04\pm0.02$ & $3.73\pm 0.02$ & $0.71\pm0.02$  & $0.62\pm0.06$ & $0.61\pm0.06$ \\
L1448N-A  & $67.5\pm10.5$  & $71.5\pm 7.6$ & $38\pm6$ & $14\pm12$ & $0.30\pm0.05$ & $0.11\pm0.14$ & $5.49\pm0.01$ & $0.50\pm0.01$ & $0.40\pm0.05$ & $0.39\pm0.05$
\enddata
\tablenotetext{a}{Velocity gradients perpendicular to outflow directions.}
\tablenotetext{b}{Velocity gradients obtained from 2D fitting.}
\tablenotetext{c}{Directions of velocity gradients from 2D fitting.}
\tablenotetext{d}{The difference between the directions of $\nabla V_{\rm 1D}$ and $\nabla V_{\rm 2D}$.  Uncertainties are obtained based on the uncertainties from both the outflow directions and $\theta_{\rm 2D}$.}
\tablenotetext{e}{Velocities at effective radius R (Table~\ref{tbl:c18o}) based on $\nabla V_{\rm 1D}$.}
\tablenotetext{f}{Ratios between rotational energy and gravitational energy, assuming $\alpha=2$.}
\tablenotetext{g}{Peak velocities from spectral fitting (Fig.~\ref{fig:vdisp}).}
\tablenotetext{h}{Total velocity dispersions from spectral fitting (Fig.~\ref{fig:vdisp}).}
\tablenotetext{i}{Approximated velocity dispersions after subtracting broadening from shifts in peak velocities ($\sigma^{2}_{\rm norot} = \sigma^{2}_{\rm tot} - \delta V_{R}^{2}$).}
\tablenotetext{j}{Non-thermal velocity dispersions ($\sigma_{\rm NT}^{2} = \sigma^{2}_{\rm norot} - \sigma^{2}_{T}$).}
\end{deluxetable*}

\section{Is Source NW Gravitationally Bound?}
\label{sect:NW}

Located at the edge of L1448N, Source NW is $\sim 3800$ AU projected from Source A (Sect.~\ref{sect:dust}), a separation close to the higher end of the binary separation distribution of protostars \citep{2013ApJ...768..110C}, and therefore it could be gravitationally unbound to the system.   
The gravitational boundness can be approximated by comparing the velocity of the source and the escape velocity. 
The escape velocity 
is calculated as $V_{e} = \sqrt{\frac{2GM_{r}}{r}}$, where $M_{r}$ is the total mass enclosed inside \r  radius $r$.  
We used $r = 3900$ AU, the distance between Source NW and the center of Sources B and A.  
By assuming a uniform density, the mass inside the radius is calculated to be $M_{r} = \frac{4}{3}\pi r^{3} \times \mu m_{\rm H} n = 1.96$ M$_{\sun}$, where $\mu=2.8$ is the mean molecular weight, $m_{\rm H}$ is the mass of hydrogen, and $n$ is the mean number density of the L1448N core.
A wide range of number densities from $\sim 3 \times 10^{5}$~cm$^{-3}$ to $4\times 10^{6}$~cm$^{-3}$ has been estimated/implied from previous work \citep{2006ApJ...638..293E,2007ApJ...668.1042K,2010ApJ...710.1247S}.
We assume $n \sim 10^{6}$~cm$^{-3}$ based on this range.  
The escape velocity is estimated to be $\sim 0.9$ km~s$^{-1}$.  

\begin{figure*}
\begin{center}
\includegraphics[scale=0.23]{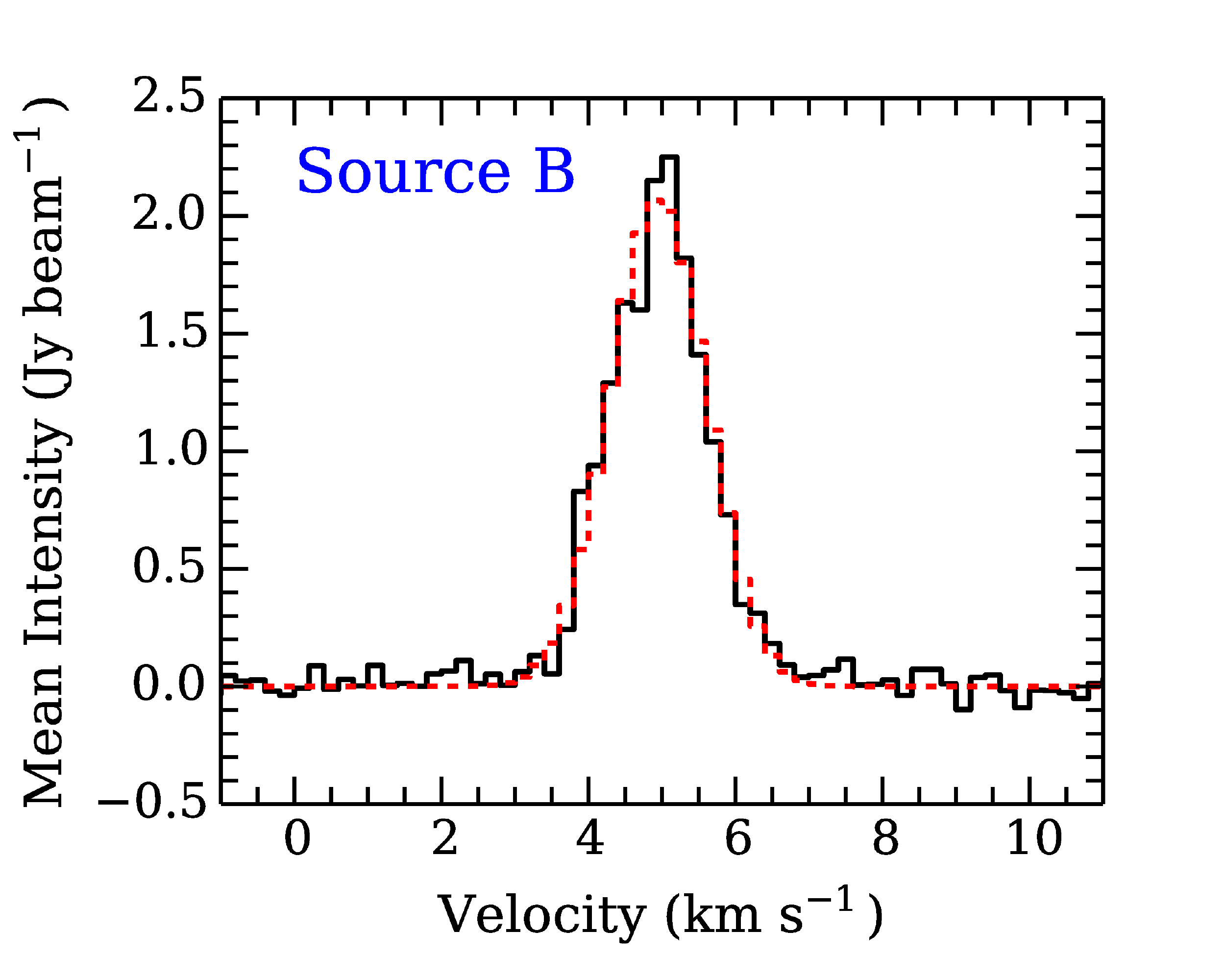}
\includegraphics[scale=0.23]{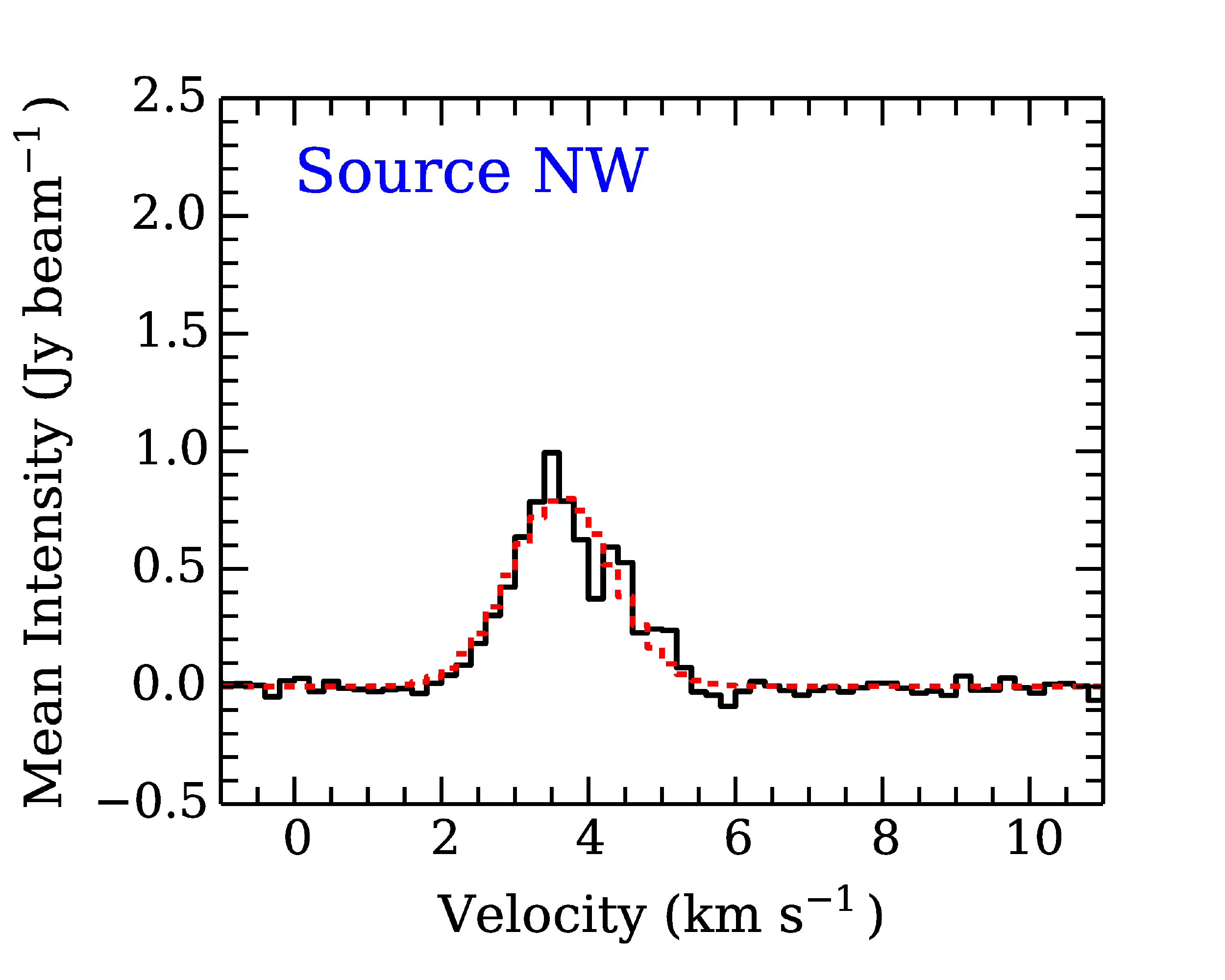}
\includegraphics[scale=0.23]{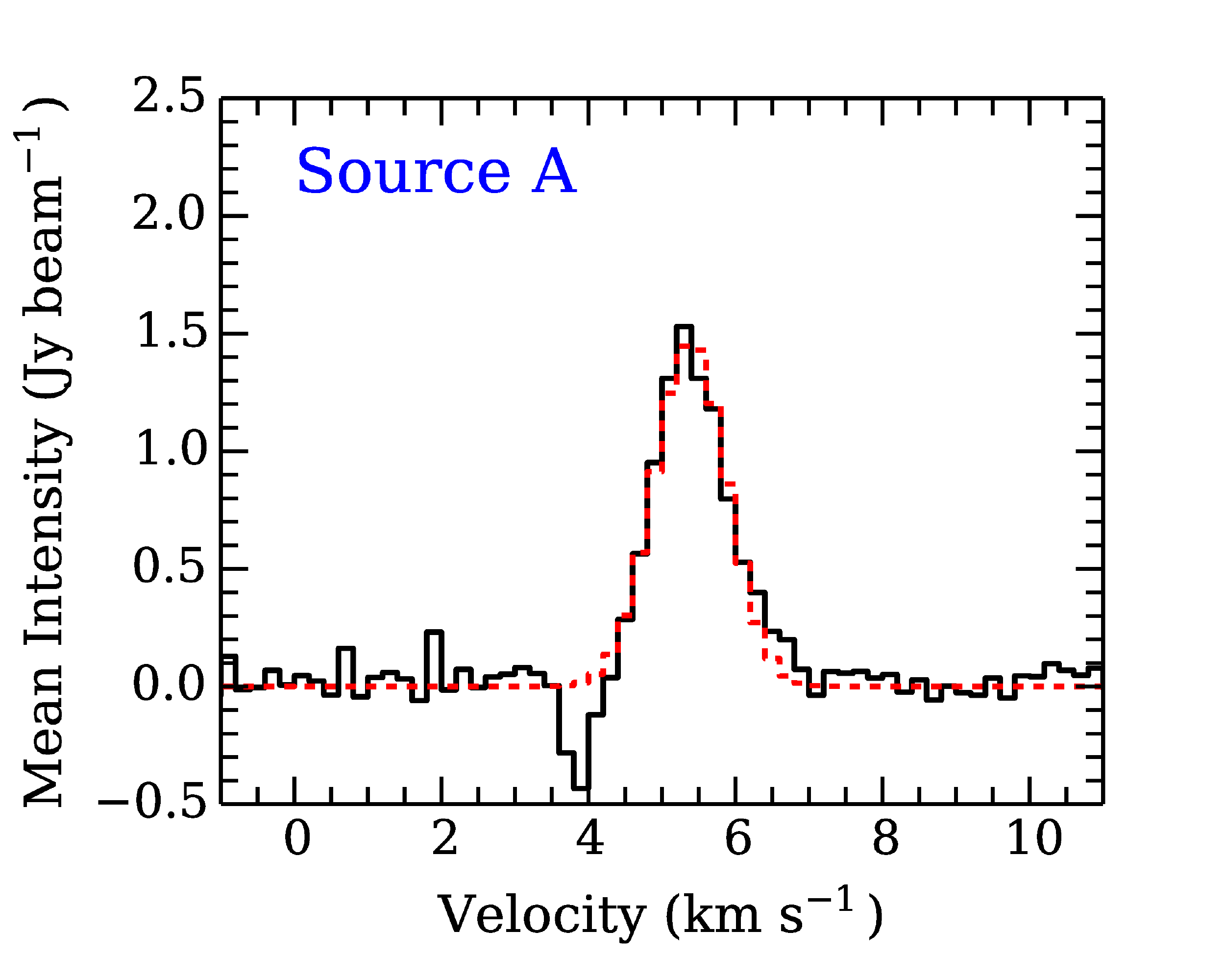}
\caption{
\footnotesize
The C$^{18}$O spectra of Sources B (left), NW (middle), and A (right).
The spectra are averaged over the FWHMs in Sources B and A (magenta lines in Fig.~\ref{fig:mom1}), and over the the whole core in Source NW.
Black lines are the observed data, and red lines are the best-fits from Gaussian fitting.
}
\label{fig:vdisp}
\end{center}
\end{figure*}

Based on the spectral fitting shown in Fig.~\ref{fig:vdisp}, the peak velocities in Sources B, A, and NW are $5.1$~km~s$^{-1}$, $5.5$~km~s$^{-1}$, and $3.7$~km~s$^{-1}$, respectively (Table~\ref{tbl:kinematics}). 
The velocity difference between Source NW and the average of Sources B and A is $\sim 1.5$~km~s$^{-1}$, larger than the escape velocity.
Therefore, Source NW is consistent with being gravitationally unbound to Sources B and A.  
This comparison assumes that the main component in the escape velocity is along the line of sight.
Also, the uncertainties from the mean density, the real separation, and the calculation of $M_{r}$ have a significant impact on the estimated escape velocity.

\section{Physical and Kinematic Environments vs.\ Multiplicity}

As discussed in Sect.~\ref{sect:dust}, L1448N is a system with three different multiplicities from source to source. 
Source B is associated with three fragments, Source NW is associated with two fragments, and Source A remains single, at the scales we can examine. 
Therefore, Sources B and NW form multiple systems, whereas Source A is a single system.  
Our C$^{18}$O observations allow us to compare the environments (including the physical and kinematic properties) and the multiplicity of the sources. 

Multiplicity in each source, assuming caused by fragmentation, occurs on a scale of 50-200 AU at the very center of each source as Figure~\ref{fig:cont} shows.
We caution that the physical and kinematic conditions that are currently observed may not necessarily be the initial conditions when fragmentation occurred at the center of each source. 
Nevertheless, the conditions at the scale of the sources ($\sim 1000$ AU), a larger scale compared to the scale of fragmentation, could still reflect the initial conditions for fragmentation due to the density differences at these two scales, which lead to differences in free-fall time-scales.
As the central region in a source has a higher density compared to the ambient environment in a whole source, the central region collapses faster in a shorter free-fall time-scale compared to the whole source.
Given the relatively short collapse time at the central region most of the initial conditions in the larger source could be preserved.
Therefore, the conditions currently observed could be reminiscent of the initial conditions that gave rise to the differing multiplicity in each source in the past. 
In addition, we assume that the \textit{relative} conditions (densities/kinematics/ages) between the three sources are the same today as they were in the past.  

We emphasize that the result is based on small number statistics; more robust conclusions will be drawn with the full datasets from MASSES.
This study serves as an illustration of the analysis that will be carried out with the complete sample from MASSES.

\subsection{Mass and Density vs.\ Multiplicity}

First, we compare the gas masses in these three sources.
The correlation between mass and the level of fragmentation in the early phases of low-mass star formation, which is expected if gravity dominates the process, is not yet clear since a few other mechanisms that complicate the correlation, such as turbulence and magnetic fields, may play critical roles.

Since the gas masses based on the 850~\um \ continuum emission are consistent with those based on the C$^{18}$O emission, we use the C$^{18}$O-derived masses for this comparison.
As Table~\ref{tbl:c18o} shows, Source B is $0.24\pm0.04$ M$_{\sun}$, Source NW is $0.28\pm0.04$ M$_{\sun}$, and Source A is $0.09\pm0.04$ M$_{\sun}$.  
The masses of Sources B and NW are comparable and are significantly larger than Source A, suggesting that more massive objects are associated with more fragmentation than the less massive one. 

Next, we compare the densities in the sources as density may be a more underlying factor than mass given the dependence between mass and density. 
We assume that these sources have uniform densities, which can be estimated as $n_{c} = M / (\frac{4}{3}\pi R^{3})$, where $M$ and $R$ are the masses and effective radii of the sources listed in Table~\ref{tbl:c18o}.
The estimated densities of Sources B and NW are $\sim 1.1 \times 10^{7}$~cm$^{-3}$ and $\sim 1.6 \times 10^{7}$~cm$^{-3}$, respectively, larger than that of Source A with $\sim 3.7 \times 10^{6}$~cm$^{-3}$.  
This suggests that the sources with multiplicity have higher densities.
Assuming that the sources have the same temperature, the Jeans masses are inversely proportional to the densities ($M_{J} \propto T^{3/2} n_{c}^{-1/2}$), and the sources with larger densities have smaller Jeans masses.  
Therefore, the sources with higher masses and smaller Jeans masses would have a higher degree of multiplicity, an outcome consistent with our observational results.
This suggests that Jeans fragmentation, controlled by gravity assuming the same thermal support, can be relevant in the fragmentation process. 
A similar conclusion was reached in the study of massive dense cores at 0.1~pc scale \citep{2015MNRAS.453.3785P} and the study of core mass function in Pipe Nebula \citep{2008ApJ...672..410L}.

\subsection{Velocity Gradient vs.\ Multiplicity}

Next, we compare the velocity gradients in these three sources.  
As Table~\ref{tbl:kinematics} shows, Source B has the largest velocity gradient, followed by Source NW, and then followed by Source A, in both $\nabla V_{\rm 1D}$ and $\nabla V_{\rm 2D}$.
This implies that the sources with multiplicity have larger velocity gradients.
It also suggests a trend between the magnitude of the velocity gradients and level of fragmentation: Source B, which has the highest number of fragments, has the largest velocity gradient among the three sources; Source A, which remains single, has the smallest velocity gradient.  

The velocity gradients are likely due to rotation since they are nearly perpendicular to the outflows (Fig.~\ref{fig:mom1}).  
Indeed, \citet{2013ApJ...772...22Y} performed radiative transfer modeling considering infall and rotation with C$^{18}$O, and suggested that rotation is the major contribution to \textit{observed velocity gradients} perpendicular to outflow directions.  
Therefore, assuming that the velocity gradients are due to rotation, $\nabla V_{\rm 1D}$ would be a better indicator for rotational velocities than $\nabla V_{\rm 2D}$.  
We caution that the above modeling considers axial symmetry with axis ratios close to spherical geometry; 
objects with filamentary geometries may have major contributions from infall motions to the observed velocity gradients along the direction perpendicular to outflows \citep{2012ApJ...748...16T,2013ApJ...772..100L,2011MNRAS.411.1354S}.

We assume that the velocity gradients are due to rotation.  
The trend between the velocity gradients and level of fragmentation is more obvious in $\nabla V_{\rm 1D}$ than in $\nabla V_{\rm 2D}$, suggesting that there is a trend between the rotational velocities and level of fragmentation. 
Combining with the result in comparing with the masses, this also suggests that more massive objects could rotate faster.  

\subsection{$\beta_{\rm rot}$ vs.\ Multiplicity}

To investigate whether or not rotation could play a role in the fragmentation process, we compare the ratio between rotational energy and gravitational energy ($\beta_{\rm rot}$): 
$\beta_{\rm rot} = \frac{1}{3}(\frac{R^{3} \Omega^{2}}{G M}) (\frac{5-2\alpha}{5-\alpha})$, where R is the radius, $\Omega$ is the angular velocity (assumed to be $\nabla V_{\rm 1D}$), $M$ is the mass of the source (we use the C$^{18}$O masses listed in Table~\ref{tbl:c18o}), and $\alpha$ is the power-law index in a density profile \citep[assumed to be 2;][]{1977ApJ...214..488S}. 
Note that $\beta_{\rm rot}$ increases with decreasing $\alpha$.  
The ratios for Sources B, NW, and A are $\sim 0.10$, $\sim 0.04$, and $\sim 0.11$, respectively. 
Therefore, we do not observe that the sources with multiplicity are associated with larger/smaller $\beta_{\rm rot}$; nor do we observe a trend between $\beta_{\rm rot}$ and the level of fragmentation. 

Numerical simulations have shown that a rotating, magnetized core could fragment with initial $\beta_{\rm rot} > 0.01$ \citep{1999ApJ...520..744B} or $\beta_{\rm rot} \geq 0.04$ \citep{2005MNRAS.362..382M}, although $\beta_{\rm rot}$ is expected to be less than $0.25-0.3$ otherwise a core would become unstable \citep[e.g.,][]{1982PThPh..68.1949H,1984ApJ...285..721T}.
All three sources have $\beta_{\rm rot} > 0.04$, consistent with the criteria for fragmentation from the simulations.  
However, it is not yet clear if there exists a correlation between $\beta_{\rm rot}$ values and the level of fragmentation for rotationally dominant fragmentation. 
More numerical simulations investigating the correlation between $\beta_{\rm rot}$ and level of fragmentation are needed to have a better understanding toward the role of rotation in the fragmentation process.

If we assume uniform density, $\beta_{\rm rot}$ for Sources B, NW, and A would be $\sim 0.3$, $\sim 0.12$, and $\sim 0.33$, respectively.  
The range of the $\beta_{\rm rot}$ values is larger than the $\beta_{\rm rot}$ in dense cores assuming uniform density (0.01-0.1) from \citet{1993ApJ...406..528G,2002ApJ...572..238C}.
The increase in $\beta_{\rm rot}$ from large scales in dense cores to small scales in protostars is consistent with equilibrium analysis considering only gravity and rotation, which expects that $\beta_{\rm rot}$ increases as a core contracts \citep{2002ARA&A..40..349T}.


\citet{2013A&A...552A.129V} has investigated the dependence of disk fragmentation on the initial core masses and $\beta_{\rm rot}$. 
According to this study, our measured $\beta_{\rm rot}$ values are consistent with the multiplicity in each source arising from disk fragmentation if the initial core masses are larger than $\sim 0.4$ M$_{\sun}$.  
The current core masses are measured to be 0.24 M$_{\sun}$ (Source B), 0.28 M$_{\sun}$ (Source NW), and 0.09 M$_{\sun}$ (Source A).  
Given that our measured masses for Sources B and NW (the two that have fragmented) are within less than a factor of two of 0.4 M$_{\sun}$, and the fact that our masses suffer from spatial filtering and do not include that mass that has already accreted onto the protostars, 
our observations are consistent but do not verify the results in \citet{2013A&A...552A.129V} for disk fragmentation.

\subsection{Non-Thermal Velocity Dispersion vs.\ Multiplicity}

Finally, we compare non-thermal dispersions in the three sources.   
As listed in Table~\ref{tbl:kinematics}, 
Sources B, NW, and A have velocity dispersions of $\sim 0.44$ km~s$^{-1}$, $\sim 0.61$ km~s$^{-1}$, and $\sim 0.39$ km~s$^{-1}$, respectively, after accounting for simple rotation and thermal broadening.  
We observe that Sources B and NW with multiplicity have larger non-thermal velocity dispersions than Source A without multiplicity.  
However, the difference between Source B and Source A is not significant considering the uncertainties.
In addition, we do not observe a trend between these non-thermal velocity dispersions and the level of fragmentation. 
This lack of trend might suggest that turbulence may not be the dominant mechanism in regulating the fragmentation process at this scale (a few hundred AU), or that turbulence played some role initially but has decayed quickly at this scale \citep[e.g.,][]{1999ApJ...524..169M}.  


\section{Summary}

In this paper we have presented SMA 230 GHz and 345 GHz observations of dust continuum and molecular lines (C$^{18}$O ($J=2\rightarrow 1$), $^{12}$CO ($J=2\rightarrow 1$), $^{13}$CO ($J=2\rightarrow 1$), N$_{2}$D$^{+}$ ($J = 3\rightarrow 2$), and HCO$^{+}$ ($J=4\rightarrow 3$) toward three protostars in Perseus (L1448N-B, L1448N-A, and L1448N-NW).  
These data are part of the large project ``Mass Assembly of Stellar Systems and their Evolution with the SMA (MASSES)", which aims to develop a more complete understanding of the stellar mass assembly process. 
The observations were performed with both the Extended and Subcompact configurations, providing angular resolutions ranging between 0.8\arcsec \ ($\sim 180$ AU) and 4.5\arcsec \ ($\sim 1000$ AU).  
By comparing the SMA data with data from the complimentary VANDAM survey at 8~mm with an angular resolution of 15 AU \citep{2015ApJ...798...61T}, 
our study has provided the first direct, comprehensive investigation on the relation between multiplicity and core kinematics in low-mass protostars.  
The main results from the study are summarized below.

1. We have detected three dust continuum sources at both 230 GHz and 345 GHz corresponding to L1448N-B (Source B), L1448N-NW (Source NW), and L1448N-A (Source A) in the literature. 
The projected separation between Source B and A is $\sim 1600$ AU, and that between Source A and Source NW is $\sim 3800$ AU.  

2. The VANDAM data have shown that Source B is associated with three 8~mm objects, Source NW is associated with two, and Source A remains single.  
The projected separations between the 8~mm objects in each of Sources B and NW range from 50 AU to 200 AU.  
Together with the three dust continuum sources separated by a few thousand AU, these results have demonstrated that L1448N is a system with hierarchical fragmentation where fragmentation occurs at various spatial scales. 

3. Three outflows have been identified in the $^{12}$CO channel maps and each outflow is associated with one continuum source. 

4. All three of the continuum sources have corresponding emission peaks in C$^{18}$O, $^{13}$CO, and HCO$^{+}$. 
The C$^{18}$O emission peaks coincide well with the three continuum sources and show extended morphologies around the continuum sources, suggesting that C$^{18}$O traces protostellar envelopes.
The $^{13}$CO and HCO$^{+}$ emission show structures beyond the three continuum sources, which possibly trace starless fragments, suggesting that L1448N harbors younger objects in addition to protostars.
N$_{2}$D$^{+}$ and C$^{18}$O emission peaks are not well coincided due to chemical destruction of N$_{2}$D$^{+}$.

5. Source NW appears to be gravitationally unbound to the rest of L1448N as suggested by the analysis of escape velocity.

6. Based on C$^{18}$O emission, the masses of three C$^{18}$O sources are estimated to be $0.24\pm 0.04$ M$_{\sun}$, $0.28\pm0.04$ M$_{\sun}$, and $0.09 \pm 0.04$ M$_{\sun}$, for Source B, Source NW, and Source A, respectively.
This result suggests that sources with multiplicity (Sources B and NW) are more massive.

7. The sources with multiplicity have higher densities.  
Assuming temperature is the same in all three sources, 
this suggests that thermal Jeans fragmentation may be relevant in the fragmentation process.

8. We have investigated velocity gradients with C$^{18}$O based on the directions perpendicular to the outflows and 2D fitting. 
In both cases, Source B has the largest magnitude, followed by Source NW, and then Source A.
The velocity gradients from 2D fitting are consistent to within 20 degrees of being perpendicular to the outflow directions.
We have observed that the sources with multiplicity have larger velocity gradients than the source without multiplicity.  
In addition, we have observed a trend between the velocity gradients and the level of fragmentation.

9. We have investigated the ratios between rotational and gravitational energy ($\beta_{\rm rot}$) in the sources.
We have not observed a difference in $\beta_{\rm rot}$ between sources with multiplicity and that without multiplicity; nor have we observed a trend between $\beta_{\rm rot}$ and the level of fragmentation.
The role for rotation in the fragmentation process is not yet clear; more simulations are needed to investigate the correlation between $\beta_{\rm rot}$ and the level of fragmentation.

10. We have investigated the non-thermal velocity dispersions of the envelope gas after accounting for simple rotation and thermal-broadening. 
Sources with multiplicity have larger non-thermal velocity dispersions in general. 
However, we do not observe a trend between the non-thermal velocity dispersions and level of fragmentation.
These results might suggest that turbulence may not be the dominating mechanism in regulating the fragmentation process at this scale (a few hundred AU), or turbulence could play a role initially but has decayed.

We emphasize that the trends presented in this paper are based merely on three sources and are not statistically significant. 
Although small number statistics, this study serves as the first step toward more statistically robust conclusions with the complete sample from MASSES in the future.

\section{Acknowledgement}

This work is based primarily on observations obtained with the Submillimeter Array, a joint project between the Smithsonian Astrophysical Observatory and the Academia Sinica Institute of Astronomy and Astrophysics and funded by the Smithsonian Institution and the Academia Sinica.  
The authors thank the SMA staff for executing these observations as part of the queue schedule, and Charlie Qi and Mark Gurwell for their technical assistance with the SMA data.

We thank the anonymous referee for valuable comments to improve the paper.
K.I.L.\ acknowledges support from NASA grant 613845.  
M.M.D.\ acknowledges support from NASA ADAP grant NNX13AE54G and from the Submillimeter Array through an SMA postdoctoral fellowship.
T.L.B.\ also acknowledges partial support from NASA ADAP grant NNX13AE54G.
E.I.V.\ acknowledges support from the Russian Ministry of Education and Science grant 3.961.2014/K.

\bibliographystyle{apj}
\bibliography{l1448}

\end{document}